\documentclass[11pt]{article}
\usepackage{Command}

\title{Polarization Formalism for Photon--Gravitational Wave Mixing Around Magnetars}

\author{Jean-Simon C\^ot\'e\email{jean-simon.cote.5@ulaval.ca} and Jean-Fran\c{c}ois Fortin\email{jean-francois.fortin@phy.ulaval.ca}}

\affiliation{
D\'epartement de Physique, de G\'enie Physique et d'Optique,\\Universit\'e Laval, Qu\'ebec, QC G1V 0A6, Canada\\
}

\abstract{%
The Gertsenshtein effect can be used to probe the stochastic gravitational wave background at high frequencies, well above the range of standard cosmological sources. In this paper, we revisit electromagnetic-gravitational wave conversion in magnetar magnetospheres using a polarization formalism that expresses the Stokes parameters in terms of conversion probabilities and phase shifts to all orders in the Gertsenshtein coupling. In doing so, we show that the adiabatic approximation commonly used in the literature is not generally justified and then derive analytical expressions for the conversion probabilities and phase shifts in two geometries where this approximation remains valid.  From these, we derive a lower bound on the stochastic gravitational wave background from the conversion of magnetar electromagnetic emission into gravitational waves, and an upper bound by requiring that the conversion of background gravitational waves into electromagnetic radiation does not exceed the observed magnetar flux in the X-ray band.  Our results demonstrate that gravitational waves generated through the Gertsenshtein conversion of magnetar electromagnetic emission produce a negligible stochastic background, as anticipated.
\vspace{15pt}
}

\date{September 2026}

\begin{document}

\maketitle

\toc


\section{Introduction}\label{SecIntro}

The advent of gravitational wave (GW) astronomy has opened a new observational window into extreme astrophysical phenomena such as black hole \cite{LIGOScientific:2016aoc} and neutron star mergers \cite{LIGOScientific:2017vwq}. While current (LIGO, Virgo, KAGRA) and next-generation (LISA, Einstein Telescope, Cosmic Explorer) interferometric detectors are primarily sensitive to the low-frequency regime (below a few thousand Hz), there are currently no dedicated high-frequency GW (HFGW) detectors.

Since no known conventional astrophysical sources are expected to produce an appreciable signal at high frequencies, HFGWs provide a promising probe of physics beyond the Standard Model (BSM).  Indeed, from the perspective of BSM physics, a large number of processes are believed to generate HFGWs. Among them, several are expected to have occurred in the early Universe, prior to recombination, providing a unique window into the physics governing the birth of the Universe. These phenomena include inflation \cite{Cook:2011hg,domcke2016probinginflationmodelsgravitational, Saikawa:2018rcs, Eggemeier:2022gyo}, preheating \cite{Khlebnikov:1997di, Easther:2006vd, Figueroa:2017vfa}, thermal plasma \cite{ Ghiglieri:2015nfa, Ringwald:2020ist, Ghiglieri:2020mhm}, early-Universe phase transitions \cite{Witten1984rs,Hogan1986dsh, Brandenburg:2021tmp, Caprini:2025trt}, and cosmic strings \cite{Blanco-Pillado:2017oxo, Auclair:2019wcv, Servant:2023tua}, among many others. In addition, some sources may still be active in the present epoch, making HFGWs a potential smoking gun for BSM physics. Notable examples include remnants of low-mass neutron star binary mergers \cite{Bauswein:2015vxa}, primordial black hole mergers \cite{Nakamura:1997sm, Sasaki:2016jop, Wang:2019kaf}, exotic compact objects \cite{Narain:2006kx, Giudice:2016zpa, Krippendorf:2018tei}, and black hole superradiance \cite{Arvanitaki:2014wva, Peng:2025zca}.

Altogether, these processes span a vast frequency range, extending from a few kHz up to THz and beyond (see reviews \cite{Caprini:2018mtu, Aggarwal:2020olq, Renzini:2022alw}), suggesting a GW background, the stochastic GW background (SGWB), analogous to the cosmic microwave background.

From conventional physics, strongly magnetized objects such as magnetars offer an alternative avenue to study HFGWs through the Gertsenshtein effect. On the one hand, photons emitted from a magnetar can convert into GWs in the magnetosphere, leading to an irreducible HFGW background and thus to a lower bound on the SGWB characteristic strain at high frequencies.  On the other hand, the HFGW background propagating through the magnetosphere of a magnetar can convert to photons, generating a flux which must not exceed the observed photon flux of the magnetar, leading to an upper bound on the SGWB characteristic strain at high frequencies.

In this work, we describe photon--GW mixing in magnetized environments using a new polarization formalism that tracks the evolution of both photon and GW Stokes parameters. This approach allows us to express the Stokes parameters, to all orders in the Gertsenshtein coupling, in terms of conversion probabilities and mixing-induced phase shifts. We then derive leading-order analytical expressions for these quantities in two relevant geometries: radially propagating rays and incident rays aligned with the magnetic moment at finite impact parameter. To our knowledge, several of these analytical results have not been obtained previously. Using these results, we derive analytical lower bounds on the characteristic strain from gravitational waves generated through Gertsenshtein conversion of radially emitted X-rays in magnetars' magnetospheres, and analytical upper bounds on the SGWB by requiring that the photon flux produced via inverse Gertsenshtein conversion does not exceed the observed X-ray flux. Applying this framework to the measured X-ray spectra of five magnetars, we derive lower and upper bounds on the SGWB characteristic strain $h_c$ in the X-ray frequency range of the order $10^{-53}\lesssim h_c\lesssim10^{-20}$.  The lower bound, which is extremely tiny, provides further support to the smoking gun signature of the BSM origin of HFGWs in the X-ray band.

\begin{figure}[t]
    \centering
    \includegraphics[width=0.95\linewidth]{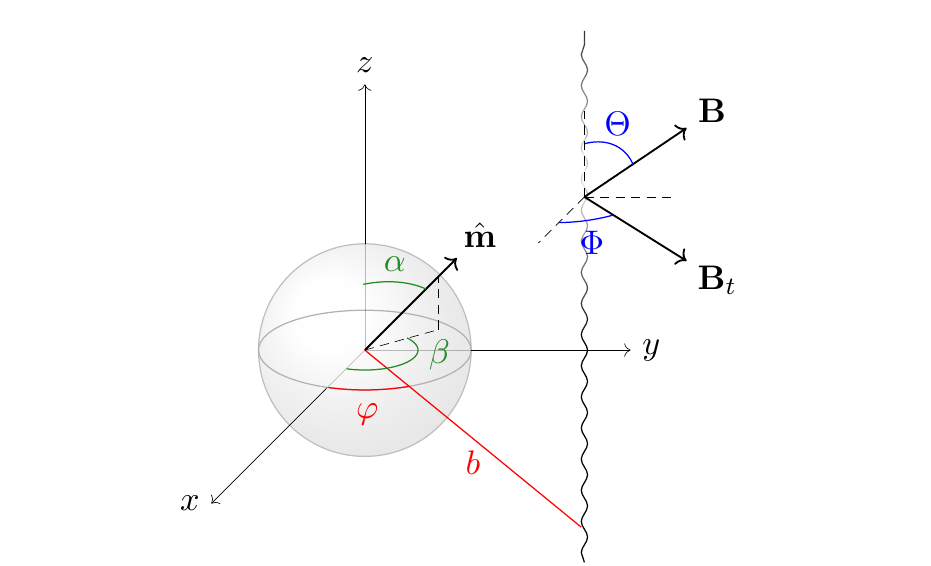}
    \caption{Geometry for the magnetosphere of a magnetar and the electromagnetic--gravitational wave propagation.  The magnetar is centered at the origin, with the magnetic field assumed to follow a dipolar configuration dictated by the magnetic moment $\mathbf{\hat{m}}$.  An electromagnetic--gravitational wave propagates along the $\mathbf{\hat{z}}$ direction with an impact parameter $b$, subject to a magnetic field $\mathbf{B}$ with transverse components $\mathbf{B}_t$.}
    \label{fig:geometry}
\end{figure}

To understand why we focus on only two geometries, we must first introduce some notation.  For a magnetar, assuming a dipolar magnetic field, the background magnetic field is given by
\begin{align}
    \mathbf{B} &= \frac{B_0}{2}\qty(\frac{r_0}{r})^3\qty[3\mathbf{\hat{r}}\qty(\mathbf{\hat{m}}\cdot \mathbf{\hat{r}}) - \mathbf{\hat{m}}], \hspace{25pt} \qty(r > r_0),\label{DipoleMagneticField}
\end{align}
where $r_0$ denotes the neutron star radius and $B_0$ the magnetic field strength at the surface, evaluated at $\mathbf{r}=r_0\mathbf{\hat{m}}$ \cite{Domcke:2025qlw}. The magnetic moment direction $\mathbf{\hat{m}} = \qty(\sin\alpha \cos \beta, \sin\alpha \sin\beta, \cos \alpha)$ is parameterized in spherical coordinates where $\alpha$ and $\beta$ denote the polar and azimuthal angles, respectively. The position vector $\mathbf{r}$ describes the trajectory of an incident wave packet propagating along the $z$-direction with impact parameter $b$ and azimuthal angle $\varphi$ in the $x$--$y$ plane. The transverse components of the magnetic field~\eqref{DipoleMagneticField} are denoted by $\mathbf{B}_t$. The angles $\Theta$ and $\Phi$ are the polar and azimuthal angles of the magnetic field, respectively. The corresponding geometry is illustrated in Figure~\ref{fig:geometry}.

With these definitions in place, we can now explain why we put the emphasis on only two geometries: waves propagating radially with a magnetic moment fixed in space; and waves propagating collinearly with the magnetic moment which is again fixed in space.  The reason concerns the spatial variation of the magnetic field azimuthal angle $\Phi$ along the direction of propagation that appears in the equations of motion after rotating to the photon perpendicular-parallel (GW plus-cross) frame.  In the literature, this quantity is usually neglected, but it is actually dominant over the photon-GW mixing in the Gertsenshtein effect for a generic geometry.  This is in contrast to the axion case where it is subdominant, owing to the much larger axion coupling constant when compared to the inverse Planck mass (the graviton coupling constant).  Hence, for the Gertsenshtein effect, the adiabatic approximation associated to the irrelevance of the variation of the magnetic field azimuthal angle $\Phi$ is not justified: the decoupling of the modes in the equations of motion occurs only when the magnetic field azimuthal angle $\Phi$ stays constant along the trajectory of the wave, restricting our analysis to the two geometries described above.

This paper is organized as follows: Section~\ref{SecPolGen} reviews the Gertsenshtein effect and solves the evolution of the Stokes parameters in terms of conversion probabilities and phase shifts.  Section~\ref{sc:geometries} describes the two relevant geometries and computes the associated conversion probabilities and phase shifts analytically.  Section~\ref{SecBounds} uses the machinery developed in the previous two sections to derive lower and upper bounds on the high-frequency SGWB.  A series of appendices discuss a generalization of the stationary phase method (Appendix~\ref{Appendix:SPMethod}), the relevant parameters of the chosen magnetars (Appendix~\ref{Appendix:MagnetarsParameters}), and one analytical result of interest (Appendix~\ref{Appendix:cte}).

Throughout this work, we adopt as a reference case the magnetar 1E~1547.0-5408, with magnetic field $B_0 = 3.2 \times 10^{14}\, \text{G}$, since this value is representative of the average magnetic field of the magnetars listed in the \href{https://www.physics.mcgill.ca/~pulsar/magnetar/main.html}{McGill Online Magnetar Catalog} \cite{Olausen:2013bpa}. We also take $r_0 = 10$ km \cite{Kaspi:2017fwg} and assume the magnetic moment to be aligned with the rotation axis \cite{amazMu2019TheFD}.


\section{Gertsenshtein Effect: General Theory}\label{SecPolGen}

In this section, we discuss the theory of the Gertsenshtein effect and derive the equations of motion for the coupled photon-GW system. We then introduce a new polarization formalism in which the evolution of the Stokes parameters is obtained by averaging over an initial relative phase. This provides a description of the mixed photon--GW Stokes parameters to all orders in the Gertsenshtein coupling constant and can be readily generalized to other photon mixing systems.

To do this, we consider a strong, non-homogeneous magnetic field, such as that of a magnetar, which justifies the use of the Euler--Heisenberg effective Lagrangian~\cite{Heisenberg:1936nmg,McDonald:2024nxj}
\begin{align}
    \mathcal{L} &= \frac{1}{2}\qty(\del^\mu h^{\rho \sigma }) \qty(\del_\mu h_{\rho \sigma})-\frac{1}{4} F^{\mu\nu}F_{\mu\nu} + \frac{1}{m_P} h_{\mu\nu}F^{\mu \rho}{F^{\nu}}_\rho +\frac{\alpha^2}{90m_e^4}\qty[\qty( F^{\mu \nu}F_{\mu\nu})^2 + \frac{7}{4}  \big( F^{\mu\nu}\tld F_{\mu\nu})^2],\label{eqnL}
\end{align}
where $h_{\mu\nu}$ is the perturbation of the metric, $F_{\mu\nu}$ is the electromagnetic field strength, $m_P\approx2.44\times10^{24}\,\text{keV}$ is the reduced Planck mass, $m_e\approx511\,\text{keV}$ is the electron mass, and $\alpha\approx1/137$ is the fine structure constant.  We then compute the evolution of both photon and graviton Stokes parameters in terms of the conversion probabilities and phase shifts induced by the magnetic field.

Our derivation follows the general oscillation formalism developed in \cite{raffelt_mixing_1988}: we work with the Euler--Heisenberg effective Lagrangian in the weak-dispersion limit and assume that the characteristic length scale over which the background magnetic field varies is much larger than the wavelength of the wave packet. We moreover restrict our analysis to high-frequency waves for which plasma effects can be safely neglected, as long as the frequencies do not invalidate the use of the Euler-Heisenberg Lagrangian. This is satisfied for frequencies
\begin{align}
      \omega_{\text{pl}}\sqrt{\frac{45\pi}{7\alpha}} \qty(\frac{B_{\text{crit}}}{B_t})\ll\omega\ll m_e, \label{FrequencyCondition}
\end{align}
where $B_t$ is the transverse magnetic field to the direction of propagation and $B_{\text{crit}}=m_e^2/\sqrt{4\pi\alpha} \approx 4.414 \times 10^{13}\, \text{G}$ is the critical magnetic field strength. The lower bound on $\omega$ in~\eqref{FrequencyCondition} coincides with the condition $\omega \gg 10^{-4}\, \text{eV}$ for a typical neutron star with $B_0 \sim 10^{14}\, \text{G}$ and $\omega_{\text{pl}}\sim10^{-5}\, \text{eV}$  \cite{raffelt_mixing_1988}.\footnote{To avoid a resonance in the plasma contribution, one should also verify that the frequency $\omega$ does not match the electron cyclotron frequency $\omega_c=m_e(B/B_{\text{crit}})$ \cite{Lai_2006, Fortin:2019npr}.  Since the plasma in a magnetar's atmosphere is of the order of centimeters \cite{Lai_2006} while the radius of a neutron star is of the order of ten kilometers, a resonance can only occur close to the surface where $B=B_0\sim10^{14}\,\text{G}$ and thus $\omega_c=m_e(B_0/B_{\text{crit}})\sim 2m_e$ which should not be a concern, leaving \eqref{FrequencyCondition} as the relevant constraint.}  As we will see in Section~\ref{SecBounds}, our bounds are safely in this window, between approximately $10\,\text{keV}$ and $200\,\text{keV}$, where the plasma effects are completely negligible even away from the magnetar surface~\cite{raffelt_mixing_1988}.

We also note that the magnetic field of some neutron stars can approach or even exceed $B_{\text{crit}}$ at their surface, in which case higher-order corrections to the Euler--Heisenberg Lagrangian should, in principle, be included \cite{Potekhin:2004jr}. Nevertheless, as the conversion takes place sufficiently far from the neutron star surface (as will be shown), we can assume that these corrections have a negligible impact on the process and that the dipolar approximation~\eqref{DipoleMagneticField} is justified.


\subsection{Equations of Motion} \label{ssc:EOM}

First, we rederive the Schrödinger-like equation obtained in \cite{raffelt_mixing_1988}. We work in the transverse-traceless gauge for the gravitational wave
\begin{align}
    \del_\mu h^{\mu\nu} &= 0, & \qty[h_{\mu\nu}] &= \frac{1}{\sqrt{2}}\mqty(0 & 0 & 0 & 0 \\0 & h_{xx} & h_{xy} & 0 \\ 0 & h_{xy} & -h_{xx} & 0 \\ 0 & 0 & 0 & 0), \label{GWgauge}
\end{align}
and in the Lorenz and temporal gauge for the photon
\begin{align}
    \del_\mu A^\mu &= 0, & A_\mu &= \qty(0, A_i). \label{Photongauge}
\end{align}
To obtain the equations of motion, we use Euler--Lagrange with the Lagrangian \eqref{eqnL} and then consider a wave packet propagating in a purely magnetic background field $\mathbf{B}$. Since the characteristic scales over which these fields vary differ significantly, we assume that they each independently satisfy their own equations of motion. The photon and the gravitational waves can then be approximately treated as plane waves propagating in a specific direction (here $z$), with $A_\mu, h_{\mu\nu} \propto e^{i(\omega t-kz)}$ and dispersion relation $\omega \approx k$. By redefining the fields with appropriate phase factors so that all quantities are real, the evolution equations can be written as \cite{Chiba:2025odu,ejlli_graviton-photon_2019, Gupta:2026scx}
\begin{align}
    \del^2\mqty(A_x \\ A_y \\ h_{xx} \\ h_{xy}) =  \mqty(\frac{4 \alpha^2 \omega^2}{45m_e^4}(7B_x^2 + 4B_y^2)  & \frac{4 \alpha^2 \omega^2}{45m_e^4}3B_xB_y & \frac{\sqrt{2}\omega}{m_P}B_y & \frac{\sqrt{2}\omega}{m_P}B_x   \\ \frac{4 \alpha^2 \omega^2}{45m_e^4}3B_xB_y & \frac{4 \alpha^2 \omega^2}{45m_e^4}(4B_x^2 + 7B_y^2) & \frac{\sqrt{2}\omega }{m_P}B_x  & -\frac{\sqrt{2}\omega }{m_P}B_y  \\
     \frac{\sqrt{2}\omega }{m_P}B_y  & \frac{\sqrt{2}\omega }{m_P}B_x & 0 & 0  \\ \frac{\sqrt{2}\omega }{m_P}B_x  & -\frac{\sqrt{2}\omega }{m_P}B_y & 0 & 0)\mqty(A_x \\ A_y \\ h_{xx} \\ h_{xy}). \label{EOM_xy}
\end{align}
Since the refractive properties of the vacuum imply $\omega\approx k$, we can simplify \eqref{EOM_xy} with the help of the WKB approximation to write $\del^2 = -\qty(\omega + i\del_z)\qty(\omega-i\del_z) \approx -2\omega \qty(\omega - i\del_z)$. We then express the background magnetic field in spherical coordinates as $\mathbf{B} = B\qty(\sin\Theta\cos \Phi, \sin\Theta\sin \Phi, \cos\Theta)$ and introduce the following change of bases
\begin{align}
    \mqty(A_\perp \\ A_\parallel \\ h_{\times} \\ h_{+}) = \mqty(\sin \Phi & -\cos\Phi & 0 & 0 \\ \cos\Phi &\sin\Phi & 0 & 0 \\ 0 & 0 & \sin 2\Phi & \cos 2\Phi \\ 0 & 0 &-\cos 2\Phi & \sin 2\Phi )\mqty(A_x \\ A_y \\ h_{xx} \\ h_{xy}). \label{PerpParBasis}
\end{align}
In this basis, the two photon components are perpendicular and parallel with respect to the transverse magnetic field $B_t = B\sin\Theta$ and the gravitational wave polarizations correspond to the plus-mode and cross-mode in this same reference frame. Following these steps and redefining the fields to obtain a symmetric quantity, \eqref{EOM_xy} takes the following Schrödinger-like form
\begin{align}
    i\del_z\mqty(A_\perp \\ h_+ \\ A_\parallel \\ h_\times) &=  \qty[\omega\Id + \mqty(\Delta_\perp & \Delta_M & \Delta_\Phi & 0 \\ \Delta_M & 0 & 0 & 2\Delta_\Phi \\ \Delta_\Phi & 0 & \Delta_\parallel &\Delta_M \\ 0 & 2\Delta_\Phi & \Delta_M & 0  )]\mqty(A_\perp \\ h_+ \\ A_\parallel \\ h_\times), \label{EOM_perpar}
\end{align}
where the multiple $\Delta$'s are
\begin{align}
    \Delta_\perp &= \frac{2\alpha \omega }{45\pi} \qty(\frac{B_t}{B_\text{crit}})^2, &   \Delta_\parallel &= \frac{7\alpha \omega }{90\pi}  \qty(\frac{B_t}{B_\text{crit}})^2, & \Delta_M &= \frac{B_t}{\sqrt{2}m_P}, & \Delta_\Phi &= \dv{\Phi}{z}. \label{DefDelta}
\end{align}
Equation~\eqref{EOM_perpar} was obtained in \cite{raffelt_mixing_1988, Liu:2023mll, Dandoy:2024oqg} except for $\Delta_\Phi$ which was assumed small and therefore neglected. This assumption is warranted in the radial geometry since $\Delta_\Phi$ is identically zero as long as the axis of rotation of the star coincides with the magnetic moment, as mentioned above and proved below. Hence this issue does not arise for the conversion of axions to photons emitted by a magnetar as long as the magnetic moment is fixed in space. In fact, for axions $\Delta_\Phi$ is several orders of magnitude smaller than $\Delta_M^{\text{axion}}$ even when the magnetic moment rotates, which justifies neglecting it. Indeed, for a magnetar period of the order of five seconds \cite{Olausen:2013bpa}, $\Delta_\Phi r_0\sim10^{-5}$ while $\Delta_M^{\text{axion}}r_0=g_aB_t/2\sim10^{-1}$ with an axion coupling constant $g_a\sim10^{-10}\,\text{GeV}^{-1}$, $B_t\sim10^{14}\,\text{G}$ and $r_0\sim10\,\text{km}$. This is however not the case for gravitons due to the weakness of the gravitational coupling constant when compared to a typical axion coupling constant. In fact, with the same magnetic field and magnetar radius, not only $\Delta_Mr_0\sim10^{-5}$ which is of the same order than $\Delta_\Phi r_0\sim10^{-5}$, but also $\Delta_Mr_0$ will drop as the third power of the distance to the magnetar, contrary to $\Delta_\Phi r_0$. Thus, for gravitational waves, the adiabatic approximation $\Delta_\Phi\sim0$ is only justified when the magnetic moment does not move, contrary to axions.

Moreover, for gravitational waves we will also be interested by conversion with non-zero impact parameters, where the adiabatic approximation $\Delta_\Phi \sim 0$ still fails unless the magnetic moment is aligned with the direction of propagation, as shown later. For example, for an incident wave packet with $\alpha = \varphi = \frac*{\pi}{2}, \beta = \frac*{\pi}{4}$ at $b=z = 100\,r_0$ in the geometry shown in Figure~\ref{fig:geometry}, $\Delta_\Phi r_0$ reaches $10^{-2}$ while the mixing parameter $\Delta_M r_0$ barely reaches $10^{-11}$ at the same point, both for $B_0 = 3.2\times10^{14}\,\text{G}$ and $r_0 = 10\,\text{km}$, invalidating the adiabatic approximation.

To circumvent this issue, we consider here two geometries (radial and parallel to the magnetic moment, which is assumed fixed in direction, see Figure~\ref{fig:geometry}) for which the azimuthal angle of the magnetic field $\Phi$ is constant along the wave-packet trajectory, so that $\Delta_\Phi = 0$ exactly. The equations of motion then decouple into two independent systems composed of $\qty{A_\perp, h_+}$ and $\qty{A_\parallel, h_\times}$, as in \cite{raffelt_mixing_1988}, which greatly simplifies the problem.


\subsection{Evolution Operator} \label{ssc:GeneralEvolutionOperator}

Setting $\Delta_\Phi = 0$ from now on, we obtain the main physical quantities relevant for describing the evolution of \eqref{EOM_perpar}. Since the perpendicular and parallel systems are now decoupled, we can work with the following simplified $2 \times 2$ systems,
\begin{align}
    i\dv{}{z} \Psi^{(j)}(z) &= \Ham^{(j)}(z)\Psi^{(j)}(z), \hspace{25pt} \qty(\text{no sum over $j$}),  \label{SimpleEOM_perpar}
\end{align}
where $\Psi^{(j)} = \qty(A_j \hspace{5pt} h_j)^\intercal$ with $A_j = \qty{A_\perp, A_\parallel}$ and $h_j = \qty{h_+, h_\times}$. Moreover, we can decompose the Hamiltonian in its free and interacting part $\Ham^{(j)} = \Ham_{0}^{(j)} + \delta \Ham^{(j)}$ as follows
\begin{align}
    \Ham_{0}^{(j)} &= \mqty(\omega + \Delta_j & 0 \\ 0 & \omega), & \delta \Ham^{(j)} &= \mqty(0 & \Delta_M \\ \Delta_M & 0 ), \label{Hamiltonians}
\end{align}
with $\Delta_j = \qty{\Delta_\perp, \Delta_\parallel}$. Since \eqref{SimpleEOM_perpar} adopts the same form as the Schrödinger equation, the solution is given by the evolution operator $\Psi^{(j)}(z) = \mathcal{U}^{(j)}(z, z_0)\Psi^{(j)}(z_0)$. Furthermore, since $\Delta_M$ is small, $\delta \Ham^{(j)}$ can be treated as a perturbation and the evolution operator can be separated in the product $\mathcal{U}^{(j)} = \mathcal{U}_0^{(j)}\mathcal{U}_{\text{int}}^{(j)}$ where $\mathcal{U}_0^{(j)}$ corresponds to the unperturbed evolution operator
\begin{align}
    \mathcal{U}_{0}^{(j)}(z, z_0) &= \exp\qty[-i \int_{z_0}^z\dd{z'}\Ham_{0}^{(j)}(z')], \label{FreeEvolutionOperator}
\end{align}
and $\mathcal{U}_{\text{int}}^{(j)}$ is given by the following Dyson series
\begin{align}
    \mathcal{U}_{\text{int}}^{(j)}(z, z_0) &= \mathcal{P}\qty{\exp[-i \int_{z_0}^z \dd{z'} \Ham_{\text{int}}^{(j)}(z')]}, \label{InteractionEvolutionOperator}
\end{align}
where $\mathcal{P}$ is the path-ordering operator and $\Ham_{\text{int}}^{(j)} = {\mathcal{U}_0^{(j)}}^\dagger \delta \Ham^{(j)} \mathcal{U}_0^{(j)}$ is the interaction Hamiltonian in the interaction picture.

Although a full analytical solution of \eqref{InteractionEvolutionOperator} is not known, the resulting matrix must belong to $U(2)$, since the Hamiltonians in \eqref{Hamiltonians} are Hermitian which implies conservation of the amplitudes in the Schrödinger-like equation \eqref{SimpleEOM_perpar}. Moreover, the interaction Hamiltonians $\Ham_{\text{int}}^{(j)}$ are traceless, which imposes $\det\,\mathcal{U}_{\text{int}}^{(j)} = 1$. It follows that $\mathcal{U}_{\text{int}}^{(j)} \in SU(2)$ and can therefore be written in the following general form
\begin{align}
    \mathcal{U}_{\text{int}}^{(j)} &= \mqty(\left[\mathcal{U}_{\text{int}}^{(j)}\right]_{11} & -\left[\mathcal{U}_{\text{int}}^{(j)}\right]_{21}^* \\
    \left[\mathcal{U}_{\text{int}}^{(j)}\right]_{21} & \left[\mathcal{U}_{\text{int}}^{(j)}\right]_{11}^* ), & \qty|\left[\mathcal{U}_{\text{int}}^{(j)}\right]_{11}|^2 + \qty|\qty[\mathcal{U}_{\text{int}}^{(j)}]_{21}|^2&=1, \label{general_SU2_matrix}
\end{align}
completely determined by three real parameters. In general, the magnitude of the off-diagonal elements yields the transition amplitudes since it quantifies the fraction of the initial amplitude converted into the other state. For a pure initial state, the argument of the diagonal element corresponds to the propagation phase accumulated by the initial photon or GW component, while the argument of the off-diagonal element gives the phase accumulated by the converted amplitude. These quantities therefore carry physical information about the polarization observables of the system, as will be shown, which motivates the relabeling of the matrix elements as
\begin{align}
    P_{h_j \to \gamma_j} &= \qty|\qty[\mathcal{U}_{\text{int}}^{(j)}]_{21}|^2, & \delta \phi_{h_j \to \gamma_j} &= \arg\qty(-\qty[\mathcal{U}_{\text{int}}^{(j)}]_{21}), & \delta \phi_{h_j} &= \arg\qty(\qty[\mathcal{U}_{\text{int}}^{(j)}]_{11}). \label{definition_probability_phaseshift}
\end{align}
Hence, the $4\times 4$ evolution operator $\mathcal{U}_{\text{int}}$ can be written exactly as $\mathcal{U}_{\text{int}} = \mathcal{U}_{\text{int}}^{(1)} \oplus \mathcal{U}_{\text{int}}^{(2)}$ where $\mathcal{U}_{\text{int}}^{(j)}$ is given by
\begin{align}
     \mathcal{U}_{\text{int}}^{(j)} &= \mqty(  \sqrt{1- P_{h_j \to \gamma_j} }e^{i \delta \phi_{h_j}}  & \sqrt{P_{h_j \to \gamma_j} }e^{-i \delta \phi_{h_j \to \gamma_j}} \\  -\sqrt{P_{h_j \to \gamma_j} }e^{i \delta \phi_{h_j \to \gamma_j}} & \sqrt{1- P_{h_j \to \gamma_j} }e^{-i \delta \phi_{h_j}} ). \label{general2X2_evolution_operator}
\end{align}
We emphasize that the preceding structure is non-perturbative and remains valid as long as the interaction Hamiltonians $\Ham_{\text{int}}^{(j)}$ are traceless and the amplitudes are conserved. At this point, we see that the impact of the weak mixing of the fields is completely encoded in the conversion probabilities and the off-diagonal/diagonal phase shifts given in \eqref{definition_probability_phaseshift}. We therefore seek for analytical approximations of these quantities in the limit of small coupling. To do this, we expand the interaction evolution operator \eqref{InteractionEvolutionOperator} to second order in perturbation theory\footnote{We note that it is necessary to go to second order to verify the probability conservation property to lowest non-trivial order in the gravitational-wave to photon conversion probability. }
\begin{align}
    \mathcal{U}_{\text{int}}^{(j)} &\approx 1 + (-i)\int_{z_0}^{z} \dd{z'}\Ham_{\text{int}}^{(j)}(z') + (-i)^2\int_{z_0}^{z} \dd{z'}\Ham_{\text{int}}^{(j)}(z')\int_{z_0}^{z'} \dd{z''}\Ham_{\text{int}}^{(j)}(z''). \label{SecondOrderUint}
\end{align}
It is then straightforward to extract the conversion probabilities and the phase shifts for the Gertsenshtein effect using \eqref{Hamiltonians} and \eqref{definition_probability_phaseshift}, obtaining, up to second order in perturbation theory,
\begin{equation}
\begin{gathered}
    P_{h_j \to \gamma_j} =  \qty|\int_{z_0}^z \dd{z'}\Delta_M(z') \exp\qty[-i\int_{z_0}^{z'}\dd{z''} \Delta_j(z'')]|^2, \\
    \delta\phi_{h_j} =\Im\qty{\int_{z_0}^z \dd{z'}\int_{z_0}^{z'}   \dd{z''}\Delta_M(z') \Delta_M(z'')\exp\qty[-i\int_{z''}^{z'}\dd{z'''} \Delta_j(z''')]}, \\
     \delta\phi_{h_j \to \gamma_j} = \arg\qty{i\int_{z_0}^z \dd{z'}\Delta_M(z') \exp\qty[-i\int_{z_0}^{z'}\dd{z''} \Delta_j(z'')]}, \\
\end{gathered}
\label{PerturbationAnalyticalApproximation}
\end{equation}
where the first two expressions are analogous to the well-known axion results of \cite{raffelt_mixing_1988}. Since no specific forms of $\Delta_j$ or $\Delta_M$ were assumed in the derivation, the results apply to any system described by a Hamiltonian of the form \eqref{Hamiltonians}.


\subsection{Stokes Parameters}

The Stokes parameters provide a convenient description of the polarization states of both photons and gravitational waves. For the photons, the Stokes parameters are naturally defined as
\eqn{
\begin{aligned}
    I_\gamma\qty(z) &= \int_0^{2\pi} \frac{\dd{\Delta \phi_{\gamma_{\perp0} h_{\times 0}}}}{2\pi} \qty[ \qty|A_\perp(z)|^2 +  \qty|A_\parallel(z)|^2  ], \\
    Q_\gamma\qty(z) &= \int_0^{2\pi} \frac{\dd{\Delta \phi_{\gamma_{\perp0} h_{\times 0}}}}{2\pi} \qty[ \qty|A_\perp(z)|^2 -  \qty|A_\parallel(z)|^2  ],  \\
    U_\gamma\qty(z) &= \int_0^{2\pi} \frac{\dd{\Delta \phi_{\gamma_{\perp0} h_{\times 0}}}}{2\pi} \qty[A_\perp(z)A_\parallel^*(z) +  A_\perp^*(z)A_\parallel(z) ], \\
    V_\gamma\qty(z) &= \int_0^{2\pi} \frac{\dd{\Delta \phi_{\gamma_{\perp0} h_{\times 0}}}}{2\pi} i\qty[A_\perp(z)A_\parallel^*(z) -  A_\perp^*(z)A_\parallel(z) ],
\end{aligned}
}[DefStokesParameters]
and analogously for the gravitational waves with $A_\perp\to h_+$ and $A_\parallel\to h_\times$. The averaging will be discussed shortly. To describe the evolution of the Stokes parameters along $z$, it is useful to introduce the density matrix $\rho$ defined as $\rho=\Psi \otimes \Psi^\dagger$ with $\Psi = \qty(A_\perp \hspace{5pt} h_+ \hspace{5pt} A_\parallel \hspace{5pt} h_\times)^\intercal$,
\begin{align}
    \rho  &=   \mqty( \qty|A_\perp|^2 & A_\perp h_+^* &  A_\perp A_\parallel^* & A_\perp h_\times^* \\ h_+A_\perp^*  & \qty|h_+|^2 &  h_+ A_\parallel^*  & h_+ h_\times^*    \\ A_\parallel A_\perp^*  &  A_\parallel h_+^* & \qty|A_\parallel|^2 & A_\parallel h_\times^* \\ h_\times A_\perp^*  & h_\times h_+^*  &   h_\times A_\parallel^*  & \qty|h_\times|^2  ).\label{density_matrix_definition_line1}
\end{align}
The evolution of the density matrix \eqref{density_matrix_definition_line1} along $z$ can be expressed in terms of the evolution operator as $\rho = \mathcal{U}\rho_0 \mathcal{U}^\dagger$ \cite{Dandoy:2024oqg}, where $\rho_0$ denotes the initial density matrix of the system.\footnote{Here the subscript $0$ indicates that the corresponding quantity is evaluated at $z = z_0$.} From this density matrix, the photon and gravitational-wave Stokes parameters defined in \eqref{DefStokesParameters} can be recovered by tracing $\rho$ with the appropriate $U(4)$ generators and averaging over the initial phase difference $\Delta \phi_{\gamma_{\perp0} h_{\times 0}} = \phi_{\gamma_{\perp 0}} - \phi_{h_{\times 0}}$. This averaging is justified by the fact that the mechanisms responsible for generating gravitational waves differ significantly from those producing typical astrophysical photons, as illustrated by the non-exhaustive list of HFGWs mentioned in the introduction. It is therefore reasonable to assume that the production processes of gravitational waves and photons are generally uncorrelated. Consequently, the initial phase difference between gravitational waves and photons in most Gertsenshtein systems can be treated as arbitrary. This naturally motivates averaging over the initial phase difference between the perpendicular photon and the cross-mode gravitational wave as in \eqref{DefStokesParameters}, the other phase difference being dependent. Indeed, although $\rho_0$ involves four fields, each with its own initial phase, it is noteworthy that it is fully characterized by only three independent phase differences, that we choose to be
\begin{align}
    \Delta \phi_{\gamma_{\perp 0} \gamma_{\parallel 0}} &=  \phi_{\gamma_{\perp 0} } -  \phi_{\gamma_{\parallel 0}}, & \Delta \phi_{h_{+0} h_{\times 0}} &= \phi_{h_{+ 0}} - \phi_{h_{\times 0}}, & \Delta \phi_{\gamma_{\perp0} h_{\times 0}} &= \phi_{\gamma_{\perp 0}} - \phi_{h_{\times 0}}, \label{phase_differences}
\end{align}
where the $\phi_0$'s are the phases of their respective fields at $z=z_0$.

Averaging over the initial phase difference $\Delta \phi_{\gamma_{\perp0} h_{\times 0}}$, which is effectively random, motivates us to describe the system's evolution using the averaged initial density matrix
\begin{align}
    \bar \rho_0 &= \int_0^{2\pi} \frac{\dd{\Delta \phi_{\gamma_{\perp0} h_{\times 0}}}}{2\pi} \rho_0 . \label{averaged_initial_matrix_density}
\end{align}
This averaging is less restrictive than assuming a purely photon or purely gravitational-wave initial state, as is often done in the literature, while still permitting a simple and tractable description of the initial conditions. In particular, all dependence on the initial state can now be expressed in terms of the initial photon and gravitational-wave Stokes parameters, since the averaging over $\Delta \phi_{\gamma_{\perp0} h_{\times 0}}$ eliminates all terms that mix the photon and gravitational-wave fields in \eqref{density_matrix_definition_line1}
\begin{align}
    \bar \rho_0 &= \frac{1}{2}\mqty( I_{\gamma 0 } + Q_{\gamma0} & 0 &  U_{\gamma 0 } - iV_{\gamma0} & 0 \\ 0  & I_{h0 } + Q_{h 0} &  0  & U_{h 0 } - iV_{h 0}    \\ U_{\gamma 0 } + iV_{\gamma0}  &  0 & I_{\gamma 0 } - Q_{\gamma0} & 0 \\ 0   &  U_{h 0 } + iV_{h 0} &   0  & I_{h0 } - Q_{h 0}  ). \label{averaged_initial_density_matrix}
\end{align}
The evolution of the averaged system is simply given by
\begin{align}
    \bar \rho &= \int_0^{2\pi} \frac{\dd{\Delta \phi_{\gamma_{\perp0} h_{\times 0}}}}{2\pi} \mathcal{U} \rho_0 \mathcal{U}^\dagger = \mathcal{U} \bar \rho_0 \mathcal{U}^\dagger, \label{evolution_mean_density_matrix}
\end{align}
since the evolution operator is independent of the initial wave-packet composition. Using \eqref{FreeEvolutionOperator} and \eqref{general2X2_evolution_operator}, the evolution operator of the four-state system can be constructed as $\mathcal{U} = \mathcal{U}^{(1)} \oplus \mathcal{U}^{(2)}$. This result and \eqref{averaged_initial_density_matrix} can then be substituted into \eqref{evolution_mean_density_matrix} to determine the evolution of the averaged density matrix $\bar \rho$ along $z$. The final expression can be written in terms of the conversion probabilities $P_{h_j \to \gamma_j}$, the mixing-induced phase shifts $\delta\phi_{h_j \to \gamma_j}$ and $\delta \phi_{h_j}$, the initial Stokes parameters, and the phase shift arising from nonlinear QED effects, defined as
\begin{align}
    \delta \phi_{\text{QED}} &= \int_{z_0}^z \dd{z'} \qty[\Delta_\perp(z') - \Delta_\parallel(z')]. \label{deltaphi_QED}
\end{align}
The resulting averaged density matrix $\bar \rho$ can then be traced with the appropriate $U(4)$ generators to extract the Stokes parameters evolution in $z$. By defining,
\begin{align}
    \Delta \phi_{h} &= \delta \phi_{h_+} - \delta \phi_{h_\times}, & \Delta \phi_{h\to \gamma}&= \delta \phi_{h_+\to \gamma_\perp} - \delta \phi_{h_\times \to \gamma_\parallel},
\end{align}
for conciseness, the photon Stokes parameters\footnote{Technically, the standard Stokes parameters are defined in terms of the electric field rather than the electromagnetic four-potential $A_\mu$, as in \eqref{DefStokesParameters}. However, within our framework, the physically normalized Stokes parameters are recovered by including an overall factor of $\omega^2$ in \eqref{DefStokesParameters}, which follows directly from the formalism presented in Subsection~\ref{ssc:EOM}. This factor disappears in \eqref{photonStokesParameter} and \eqref{gravitonStokesParameter}.} then take the following form
\eqn{
\begin{aligned}
    I_{\gamma} &= I_{\gamma 0 } + \qty(\frac{I_{h 0} + Q_{h 0}}{2}-\frac{I_{\gamma 0} + Q_{\gamma 0}}{2})P_{h_+ \to \gamma_\perp} + \qty(\frac{I_{h 0} - Q_{h 0}}{2}-\frac{I_{\gamma 0} - Q_{\gamma 0}}{2})P_{h_\times \to \gamma_\parallel},  \\
    Q_{\gamma} &= Q_{\gamma 0 } + \qty(\frac{I_{h 0} + Q_{h 0}}{2}-\frac{I_{\gamma 0} + Q_{\gamma 0}}{2})P_{h_+ \to \gamma_\perp} - \qty(\frac{I_{h 0} - Q_{h 0}}{2}-\frac{I_{\gamma 0} - Q_{\gamma 0}}{2})P_{h_\times \to \gamma_\parallel},  \\
    U_{\gamma} &= \qty[U_{\gamma 0} \cos\qty(\delta \phi_{\text{QED}} -\Delta \phi_{h} ) - V_{\gamma 0} \sin\qty(\delta \phi_{\text{QED}} -\Delta \phi_{h})]\sqrt{(1-P_{h_+ \to \gamma_\perp})(1-P_{h_\times \to \gamma_\parallel})}  \\
    &\hspace{50pt} + \qty[U_{h 0} \cos\qty(\delta \phi_{\text{QED}} -\Delta \phi_{h\to \gamma} ) - V_{h 0} \sin\qty(\delta \phi_{\text{QED}} -\Delta \phi_{h\to \gamma})]\sqrt{P_{h_+ \to \gamma_\perp}P_{h_\times \to \gamma_\parallel}}, \\
    V_{\gamma} &= \qty[V_{\gamma 0} \cos\qty(\delta \phi_{\text{QED}} -\Delta \phi_{h} ) +U_{\gamma 0} \sin\qty(\delta \phi_{\text{QED}} -\Delta \phi_{h} )]\sqrt{(1-P_{h_+ \to \gamma_\perp})(1-P_{h_\times \to \gamma_\parallel})}  \\
    &\hspace{50pt} + \qty[V_{h 0} \cos\qty(\delta \phi_{\text{QED}} -\Delta \phi_{h\to \gamma} ) + U_{h 0} \sin\qty(\delta \phi_{\text{QED}} -\Delta \phi_{h\to \gamma})]\sqrt{P_{h_+ \to \gamma_\perp}P_{h_\times \to \gamma_\parallel}}.
\end{aligned}}[photonStokesParameter]
For the gravitational wave, the Stokes parameters are given by
\eqn{
\begin{aligned}
    I_{h} &= I_{h 0 } - \qty(\frac{I_{h 0} + Q_{h 0}}{2}-\frac{I_{\gamma 0} + Q_{\gamma 0}}{2})P_{h_+ \to \gamma_\perp} - \qty(\frac{I_{h 0} - Q_{h 0}}{2}-\frac{I_{\gamma 0} - Q_{\gamma 0}}{2})P_{h_\times \to \gamma_\parallel},\\
    Q_{h} &= Q_{h 0 } - \qty(\frac{I_{h 0} + Q_{h 0}}{2}-\frac{I_{\gamma 0} + Q_{\gamma 0}}{2})P_{h_+ \to \gamma_\perp} + \qty(\frac{I_{h 0} - Q_{h 0}}{2}-\frac{I_{\gamma 0} - Q_{\gamma 0}}{2})P_{h_\times \to \gamma_\parallel}, \\
    U_{h} &= \qty[U_{h 0} \cos\Delta \phi_{h}  - V_{h 0} \sin \Delta \phi_{h} ]\sqrt{(1-P_{h_+ \to \gamma_\perp})(1-P_{h_\times \to \gamma_\parallel})}  \\
    &\hspace{130pt} + \qty[U_{\gamma 0} \cos\Delta \phi_{h\to \gamma} - V_{\gamma 0} \sin \Delta \phi_{h\to \gamma}]\sqrt{P_{h_+ \to \gamma_\perp}P_{h_\times \to \gamma_\parallel}}, \\
    V_{h} &= \qty[V_{h0} \cos\Delta \phi_{h} + U_{h 0} \sin \Delta \phi_{h} ]\sqrt{(1-P_{h_+ \to \gamma_\perp})(1-P_{h_\times \to \gamma_\parallel})}  \\
    &\hspace{130pt} + \qty[V_{\gamma 0} \cos\Delta \phi_{h\to \gamma} + U_{\gamma 0} \sin\Delta \phi_{h\to \gamma}]\sqrt{P_{h_+ \to \gamma_\perp}P_{h_\times \to \gamma_\parallel}}.
\end{aligned}}[gravitonStokesParameter]
We emphasize that the subscript $0$ denotes evaluation at the initial condition $z = z_0$. Since the evolution operator $\mathcal{U}_{\text{int}}$ is defined through \eqref{general2X2_evolution_operator} to all orders in perturbation theory, the expressions \eqref{photonStokesParameter} and \eqref{gravitonStokesParameter} are likewise exact to all orders. The main difficulty lies in obtaining analytic expressions for the conversion probabilities and phase shifts beyond leading order.

Nevertheless, these quantities can be computed numerically by evolving \eqref{SimpleEOM_perpar}, after which the photon and gravitational-wave Stokes parameters follow directly from the above relations, provided the initial state is averaged and expressed as in \eqref{averaged_initial_density_matrix}. The results derived here are analogous to those obtained in the axion case \cite{Fortin:2023jlg}, but the formalism presented in this section offers a more systematic and readily generalizable framework for other mixing systems.


\subsection{Astrophysical Stokes Parameters} \label{ssc:Astrophysical_Stokes_Parameters}

In the decoupling limit of gravity, the off-diagonal mixing in the interaction term of the Hamiltonians \eqref{Hamiltonians}, which arises solely from the Gertsenshtein effect, vanishes ($\Delta_M = 0$) and the evolution operator reduces to $\mathcal{U} = \mathcal{U}_0$. In this case, the difference between the initial and final states is entirely due to QED effects, namely $\Delta_\perp$ and $\Delta_\parallel$. We denote this scenario by ${\bar \rho}_{\text{a}}$, as it accounts only for the ``usual'' astrophysical processes
\begin{align}
    {\bar \rho}_{\text{a}} &= \left. \bar \rho\right|_{\Delta_M = 0} = \mathcal{U}_0 \bar \rho_0 \mathcal{U}_0^\dagger. \label{DefrhoAstro}
\end{align}
Since $\bar \rho_{\text{a}}$ is constructed from the Stokes parameters that would be observed in the absence of mixing, we refer to these as the astrophysical Stokes parameters
\begin{align}
    {\bar \rho}_{\text{a}} &\equiv \mqty( \frac{1}{2}\qty(I_{\gamma\text{a}} + Q_{\gamma\text{a}}) & 0 &  \frac{1}{2}\qty(U_{\gamma \text{a}} - iV_{\gamma \text{a}}) & 0 \\ 0  & \frac{1}{2}\qty(I_{h \text{a}} + Q_{h \text{a}}) &  0  & \frac{1}{2}\qty(U_{h \text{a}} - iV_{h \text{a}})    \\ \frac{1}{2}\qty(U_{\gamma  \text{a}} + iV_{\gamma \text{a}})  &  0 & \frac{1}{2}\qty(I_{\gamma \text{a}} - Q_{\gamma \text{a}}) & 0 \\ 0   &  \frac{1}{2}\qty(U_{h \text{a}} + iV_{h \text{a}}) &   0  & \frac{1}{2}\qty(I_{h \text{a}} - Q_{h \text{a}})  ). \label{Matrixrhoastro}
\end{align}
By substituting $\bar \rho_0$ in \eqref{evolution_mean_density_matrix} with \eqref{DefrhoAstro}, the evolution of the averaged system can equivalently be expressed in terms of the astrophysical Stokes parameters
\begin{align*}
    \bar \rho &= \mathcal{U} \mathcal{U}_0^\dagger  {\bar \rho}_{\text{a}} \mathcal{U}_0 \mathcal{U}^\dagger.
\end{align*}
As before, the Stokes parameters can be obtained by computing the evolution operators using \eqref{FreeEvolutionOperator} and \eqref{general2X2_evolution_operator}, and then substituting the result into the preceding equation. By tracing the resulting averaged density matrix with the appropriate $U(4)$ generator, one obtains the evolution along $z$ of the Stokes parameters in terms of the astrophysical Stokes parameters, the conversion probabilities, and the various phase shifts defined previously. The photon Stokes parameters then take the following form
\eqn{
\begin{aligned}
    I_{\gamma} &= I_{\gamma\text{a}} + \qty(\frac{I_{h\text{a}} + Q_{h\text{a}}}{2}-\frac{I_{\gamma\text{a}} + Q_{\gamma\text{a}}}{2})P_{h_+ \to \gamma_\perp}  + \qty(\frac{I_{h\text{a}} - Q_{h\text{a}}}{2}-\frac{I_{\gamma \text{a}} - Q_{\gamma \text{a}}}{2})P_{h_\times \to \gamma_\parallel},  \\
    Q_{\gamma} &= Q_{\gamma\text{a}} + \qty(\frac{I_{h\text{a}} + Q_{h\text{a}}}{2}-\frac{I_{\gamma\text{a}} + Q_{\gamma\text{a}}}{2})P_{h_+ \to \gamma_\perp}  - \qty(\frac{I_{h \text{a}} - Q_{h \text{a}}}{2}-\frac{I_{\gamma \text{a}} - Q_{\gamma \text{a}}}{2})P_{h_\times \to \gamma_\parallel},  \\
    U_{\gamma} &= \qty[U_{\gamma \text{a}} \cos\Delta \phi_{h}  + V_{\gamma \text{a}} \sin\Delta \phi_{h} ]\sqrt{(1-P_{h_+ \to \gamma_\perp})(1-P_{h_\times \to \gamma_\parallel})}  \\
    &\hspace{25pt} + \qty[U_{h \text{a}} \cos\qty(\delta \phi_{\text{QED}} -\Delta \phi_{h\to \gamma} ) - V_{h \text{a}} \sin\qty(\delta \phi_{\text{QED}} -\Delta \phi_{h\to \gamma} )]\sqrt{P_{h_+ \to \gamma_\perp}P_{h_\times \to \gamma_\parallel}},  \\
    V_{\gamma} &= \qty[V_{\gamma \text{a}} \cos\Delta \phi_{h}  -U_{\gamma \text{a}} \sin\Delta \phi_{h} ]\sqrt{(1-P_{h_+ \to \gamma_\perp})(1-P_{h_\times \to \gamma_\parallel})}  \\
    &\hspace{25pt} + \qty[V_{h \text{a}} \cos\qty(\delta \phi_{\text{QED}} -\Delta \phi_{h\to \gamma} ) + U_{h \text{a}} \sin\qty(\delta \phi_{\text{QED}} -\Delta \phi_{h\to \gamma} )]\sqrt{P_{h_+ \to \gamma_\perp}P_{h_\times \to \gamma_\parallel}}.
\end{aligned}
}[photonStokesParameterAstro]
For the gravitational waves, the Stokes parameters are given by
\eqn{
\begin{aligned}
    I_{h} &= I_{h \text{a}} - \qty(\frac{I_{h \text{a}} + Q_{h \text{a}}}{2}-\frac{I_{\gamma \text{a}} + Q_{\gamma \text{a}}}{2})P_{h_+ \to \gamma_\perp}  - \qty(\frac{I_{h \text{a}} - Q_{h \text{a}}}{2}-\frac{I_{\gamma \text{a}} - Q_{\gamma \text{a}}}{2})P_{h_\times \to \gamma_\parallel},  \\
    Q_{h} &= Q_{h \text{a}} - \qty(\frac{I_{h \text{a}} + Q_{h \text{a}}}{2}-\frac{I_{\gamma \text{a}} + Q_{\gamma \text{a}}}{2})P_{h_+ \to \gamma_\perp}   + \qty(\frac{I_{h \text{a}} - Q_{h \text{a}}}{2}-\frac{I_{\gamma \text{a}} - Q_{\gamma \text{a}}}{2})P_{h_\times \to \gamma_\parallel},  \\
    U_{h} &= \qty[U_{h \text{a}} \cos\Delta \phi_{h}  - V_{h \text{a}} \sin\Delta \phi_{h} ]\sqrt{(1-P_{h_+ \to \gamma_\perp})(1-P_{h_\times \to \gamma_\parallel})}  \\
    &\hspace{25pt} + \qty[U_{\gamma \text{a}} \cos\qty(\delta\phi_{\text{QED}}-\Delta \phi_{h\to \gamma} ) + V_{\gamma \text{a}} \sin\qty(\delta\phi_{\text{QED}}-\Delta \phi_{h\to \gamma} )]\sqrt{P_{h_+ \to \gamma_\perp}P_{h_\times \to \gamma_\parallel}},  \\
    V_{h} &= \qty[V_{h \text{a}} \cos\Delta \phi_{h}  + U_{h \text{a}} \sin\Delta \phi_{h} ]\sqrt{(1-P_{h_+ \to \gamma_\perp})(1-P_{h_\times \to \gamma_\parallel})}  \\
    &\hspace{25pt} + \qty[V_{\gamma \text{a}} \cos\qty(\delta\phi_{\text{QED}}-\Delta \phi_{h\to \gamma} ) - U_{\gamma \text{a}} \sin\qty(\delta\phi_{\text{QED}}-\Delta \phi_{h\to \gamma} )]\sqrt{P_{h_+ \to \gamma_\perp}P_{h_\times \to \gamma_\parallel}}.
\end{aligned}
}[gravitonStokesParameterAstro]
The Stokes parameters \eqref{photonStokesParameterAstro} and \eqref{gravitonStokesParameterAstro} provide a convenient way to incorporate gravitational-wave mixing into existing calculations that neglect off-diagonal graviton couplings. In particular, one can account for this effect by inserting the conversion probabilities and mixing-induced phase shifts derived in Section~\ref{sc:geometries}.

Within this framework, we remain agnostic about the specific astrophysical contributions encoded in $\Delta_j$ in \eqref{Hamiltonians}. Although we focus on the Euler--Heisenberg QED correction, $\Delta_j$ may also include additional effects, such as plasma contributions in the high-magnetization limit. These can be incorporated straightforwardly by substituting the corresponding astrophysical Stokes parameters into the above expressions, thereby extending them to include Gertsenshtein conversion.


\section{Geometries}\label{sc:geometries}

In this section, we seek an analytical understanding of the conversion process near a neutron star. The fully general case, with arbitrary magnetic moment orientation, is complicated. Besides greatly complicating the integrals needed to obtain the conversion probabilities and phase shifts in \eqref{PerturbationAnalyticalApproximation}, it also requires retaining the $\Delta_\Phi$ term in the equations of motion \eqref{EOM_perpar}, since it is not negligible when compared to $\Delta_M$, as mentioned earlier. Consequently, the equations of motion for $\qty{A_\perp, h_+}$ and $\qty{A_\parallel, h_\times}$ no longer decouple, which significantly increases the complexity of the problem. The generic configuration is therefore better suited to a numerical treatment and will not be addressed analytically here.

Instead, we consider two simpler cases: a radially propagating wave and an incident wave parallel to the magnetic moment, always with $\mathbf{\hat{m}}$ fixed. In both situations, the mixing term $\Delta_\Phi = 0$, since the azimuthal angle of the magnetic field $\Phi$ remains constant along the entire trajectory, as will be shown. We will compute the conversion probabilities and the off-diagonal/diagonal phase shifts defined in \eqref{PerturbationAnalyticalApproximation} and then obtain analytical approximations for these expressions. To our knowledge, the analytical expressions for the conversion probabilities in the parallel configuration and for all of the phase shifts in any of the analyzed geometries have not been obtained previously for the photon--GW system.


\subsection{Radial Case}

For a radial wave propagating from the surface of the neutron star at $z_0 = r_0$ to a distant observer at $z \to \infty$, we have $b = 0$, and the magnetic field vector \eqref{DipoleMagneticField} simplifies to
\begin{align}
   \left.\mathbf{B}\right|_{b=0} &= \frac{B_0}{2}\qty(\frac{r_0}{z})^3\qty( -\sin\alpha \cos \beta, -\sin\alpha \sin \beta, 2\cos\alpha ). \label{radial_Bfield}
\end{align}
Here, we observe that $\tan\Phi = \frac*{B_y}{B_x}= \tan\beta$ is constant, so $\Delta_\Phi = 0$ exactly. In fact, the angle $\Phi$ lies in the quadrant opposite to $\beta$. Consequently, they differ by $\pi$, and we can eliminate $\Phi$ by substituting $\Phi = \beta - \pi$ in the change of bases \eqref{PerpParBasis}. From \eqref{radial_Bfield}, one can compute the transverse magnetic field $B_t$ and the $\Delta$'s of the Hamiltonians \eqref{Hamiltonians} then correspond to
\begingroup\makeatletter\def\f@size{10}\check@mathfonts\def\maketag@@@#1{\hbox{\m@th\large\normalfont#1}}
\begin{align}
    \left. \Delta_\perp\right|_{b=0} &= \frac{\alpha \omega }{90\pi} \qty(\frac{B_0 \sin \alpha}{B_\text{crit}})^2 \qty(\frac{r_0}{z})^6, & \hspace{-1pt}\left.\Delta_\parallel\right|_{b=0} &= \frac{7\alpha \omega }{360\pi} \qty(\frac{B_0\sin\alpha}{B_\text{crit}})^2 \qty(\frac{r_0}{z})^6, & \hspace{-1pt}\left. \Delta_M\right|_{b=0}  &= \frac{B_0\sin\alpha}{2\sqrt{2}m_P}\qty(\frac{r_0}{z})^3, \label{LongDelta}
\end{align}
\endgroup
which can be abbreviated as
\begin{align}
    \left.\Delta_{j}\right|_{b = 0} &=  \Delta_{j0}\sin^2\alpha\qty(\frac{r_0}{z})^6, & \left.\Delta_M\right|_{b = 0} &=  \Delta_{M0}\sin\alpha\qty(\frac{r_0}{z})^3, \label{RadialDelta_Definition}
\end{align}
where $\Delta_{j0} = \qty{\Delta_{\perp 0 }, \Delta_{\parallel 0 }}$ and the subscript $0$ indicates that the $\Delta$'s in \eqref{LongDelta} are evaluated at $z = r_0$ and $\alpha = \frac*{\pi}{2}$.

\subsubsection{Conversion Probabilities and Phase Shifts}

The conversion probabilities and the off-diagonal phase shifts are both related to the following integral through \eqref{definition_probability_phaseshift}
\begin{align}
    \left. \qty[\mathcal{U}_{\text{int}}]_{21}\right|_{b=0} &= -i\Delta_{M0}r_0^3\sin\alpha   \int_{z_0}^z \dd{z'}\frac{1}{{z'}^3} \exp\qty(-i\Delta_{j 0}r_0^6 \sin^2\alpha\int_{z_0}^{z'} \dd{z''} \frac{1}{{z''}^6}). \label{NumericalProbInt}
\end{align}
Changing variables to $t' = \frac{\Delta_{j 0}r_0 \sin^2\alpha}{5} \frac{r_0^5}{{z'}^5}$, and taking the initial condition at the neutron star surface $z_0 = r_0$ with the observer located far away at $z\to \infty$, the integral simplifies to the subtraction of a complete and an incomplete gamma function, yielding
\eqn{
\begin{gathered}
    \left.P_{h_j\to \gamma_j}\right|_{b=0} = \frac{(\Delta_{M0}r_0)^2(\sin\alpha)^{\frac{2}{5}} }{25\qty(\frac*{\Delta_{j 0 } r_0}{5})^{\frac{4}{5}}} \qty|\Gamma\qty(\frac{2}{5}) - \Gamma\qty(\frac{2}{5}, -\frac{i \Delta_{j 0}r_0\sin^2\alpha}{5})|^2,  \\
    \left.\delta \phi_{h_j\to \gamma_j}\right|_{b=0} = \arg\qty{\qty[\Gamma\qty(\frac{2}{5}) - \Gamma\qty(\frac{2}{5}, -\frac{i \Delta_{j 0}r_0\sin^2\alpha}{5})]e^{-\frac{i\Delta_{j 0}r_0\sin^2\alpha}{5} + \frac{7i\pi}{10}}}.
\end{gathered}
}[MixingProbShiftRadialExact]
Similar results for the conversion probabilities have been obtained in \cite{Dessert:2019sgw, Fortin:2018aom, Liu:2023mll, Dandoy:2024oqg}. To derive the diagonal phase shift, we substitute \eqref{RadialDelta_Definition} into \eqref{PerturbationAnalyticalApproximation}, yielding
\begin{align}
    \left.\delta \phi_{h_j}\right|_{b=0} &= \Im\qty[\Delta_{M0}^2r_0^6\sin^2\alpha  \int_{z_0}^z \dd{z'}\int_{z_0}^{z'}  \dd{z''}\frac{1}{{z'}^3} \frac{1}{{z''}^3} \exp\qty(-i\Delta_{j 0 } r_0^6 \sin^2\alpha \int_{z''}^{z'}\dd{z'''} \frac{}{{z'''}^6} )]. \label{diagonal_phase_shift_radial_start}
\end{align}
This integral can be simplified by the change of variables $z'=r_0\tan\vartheta'$ and analogously for $z''$.  After applying the boundary conditions $z_0=r_0$ and $z\to \infty$, and reversing the integration order, the diagonal phase shift can be rewritten as
\begingroup\makeatletter\def\f@size{10}\check@mathfonts\def\maketag@@@#1{\hbox{\m@th\large\normalfont#1}}
\begin{equation}
\begin{aligned}
    \left.\delta \phi_{h_j}\right|_{b=0}  &=  \\
    &\hspace{-25pt} (\Delta_{M0}r_0\sin \alpha)^2 \int_{\frac{\pi}{2}}^{\frac{\pi}{4}}\dd{\vartheta'} \int_{\frac{\pi}{2}}^{\vartheta'} \dd{\vartheta''} \cot\vartheta'\csc^2\vartheta'\cot\vartheta''\csc^2\vartheta'' \sin\qty(\Delta_{j0}r_0 \sin^2\alpha  \int_{\vartheta''}^{\vartheta'} \dd{\vartheta'''}\cot^4\vartheta'''\csc^2\vartheta''' ).
\end{aligned}
\label{PhotonDiagPhaseshiftExactRadial}
\end{equation}
\endgroup
For completeness, we also provide the phase shift induced by nonlinear QED effects, $\delta \phi_{\text{QED}}$. Using \eqref{deltaphi_QED} together with \eqref{LongDelta}, one straightforwardly obtains
\begin{align*}
    \left.\delta \phi_{\text{QED}}\right|_{b=0}  &=  \frac{1}{5}(\Delta_{\perp 0}r_0 - \Delta_{\parallel 0}r_0)\sin^2\alpha=  -\frac{\alpha \omega r_0}{600 \pi}\qty(\frac{B_0}{B_{\text{crit}}})^2\sin^2\alpha.
\end{align*}

\subsubsection{Analytical Approximations} \label{AnalyticalApproximationsRadial}

The dimensionless quantities $\Delta_{j0} r_0$ can be quite large for typical neutron star parameters. Indeed, using $B_0 = 3.2 \times 10^{14}$ G, $r_0 = 10$ km, and a wave-packet frequency $\omega = 10$ keV corresponding to X-rays passing near the magnetar 1E~1547.0-5408, we obtain $\Delta_{j0} r_0 \sim 10^{12}$.

This motivates working in that regime, in which the quantities introduced above admit simple asymptotic forms. In particular, the conversion probabilities and the off-diagonal phase shifts can be obtained by expanding the incomplete Gamma function in \eqref{MixingProbShiftRadialExact} for large arguments and retaining the leading terms. To first order, this gives
\eqn{
\begin{gathered}
    \left.P_{h_j \to \gamma_j}\right|_{b=0} \approx \frac{\Gamma\qty(\frac{2}{5})^2(\Delta_{M0}r_0)^2 \sin^2\alpha}{25\lambda^{\frac{4}{5}}}\qty{1 + \mathcal{O}\qty[\lambda^{-\frac{3}{5}}]}, \\
    \left.\delta \phi_{h_j\to \gamma_j}\right|_{b=0}  \approx -\lambda +  \frac{7\pi}{10} + \mathcal{O}\qty[\lambda^{-\frac{3}{5}}],
\end{gathered}
}[ApproxProbShiftRadial]
where $\lambda$ is the large expansion parameter defined in \eqref{DefinitionfgthetalambdaRadial}.
For the diagonal phase shift, the stationary phase method for degenerate critical points can be used to approximate \eqref{diagonal_phase_shift_radial_start} in the limit $\lambda \gg 1$. Following the procedure outlined in Appendix~\ref{Appendix:SPMethod} with
\begin{align}
    f(\vartheta) &= 5\int_{\frac{\pi}{4}}^{\vartheta} \dd{\vartheta'} \cot^4\vartheta' \csc^2\vartheta', & g(\vartheta) &=  \cot\vartheta\csc^2\vartheta, & \lambda &= \frac{\Delta_{j0}r_0 \sin^2\alpha}{5}, \label{DefinitionfgthetalambdaRadial}
\end{align}
we apply \eqref{GeneralStationnaryPhase} to evaluate the contribution from the critical point $\vartheta_* = \frac*{\pi}{2}$ of $f(\vartheta)$, while the remaining boundary contribution away from the critical point is obtained by integration by parts. This procedure yields the following approximation:
\eqn{
\begin{aligned}
    \left.\delta \phi_{h_j}\right|_{b=0} &\approx (\Delta_{M0}r_0 \sin \alpha)^2\left\{-\frac{\sqrt{5+2 \sqrt{5}}\,\Gamma\qty(\frac{2}{5})^2 }{50 \lambda^{\frac{4}{5}}}+\frac{1}{5\lambda}+\frac{\Gamma \qty(\frac{2}{5}) \cos \qty(\lambda - \frac{\pi}{5})}{25 \lambda^{ \frac{7}{5}}} + \mathcal{O}\qty[\lambda^{-\frac{8}{5}}]\right\}.
\end{aligned}
}[RadialApproxPhaseShift]
\begin{figure}[t!]
    \begin{subfigure}{0.515\textwidth}
        \centering
        \includegraphics[width=\linewidth]{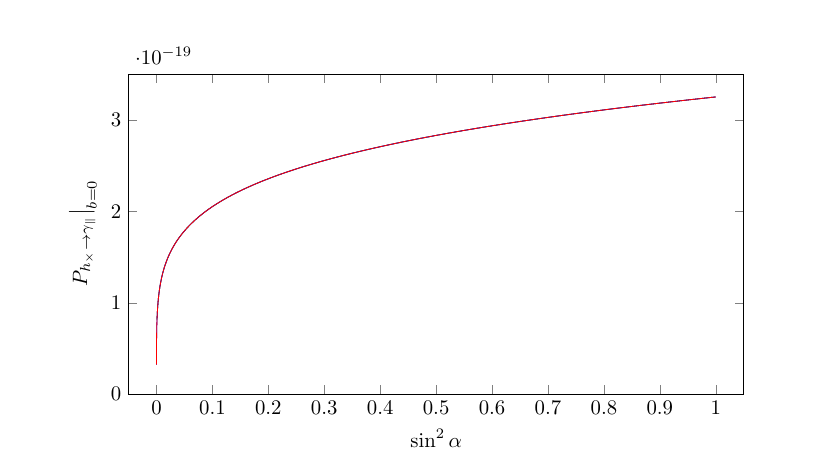}
    \end{subfigure}
    \hspace{-20pt}
    \begin{subfigure}{0.515\textwidth}
        \centering
        \includegraphics[width=\linewidth]{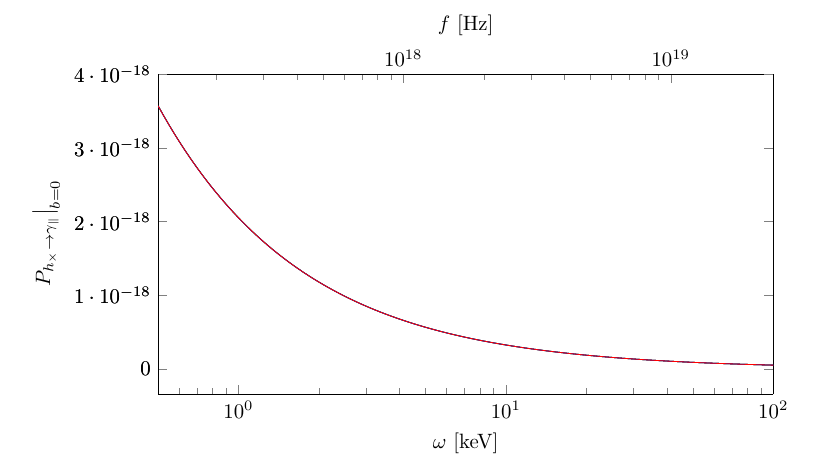}
    \end{subfigure}\\
    \begin{subfigure}{0.515\textwidth}
        \centering
        \includegraphics[width=\linewidth]{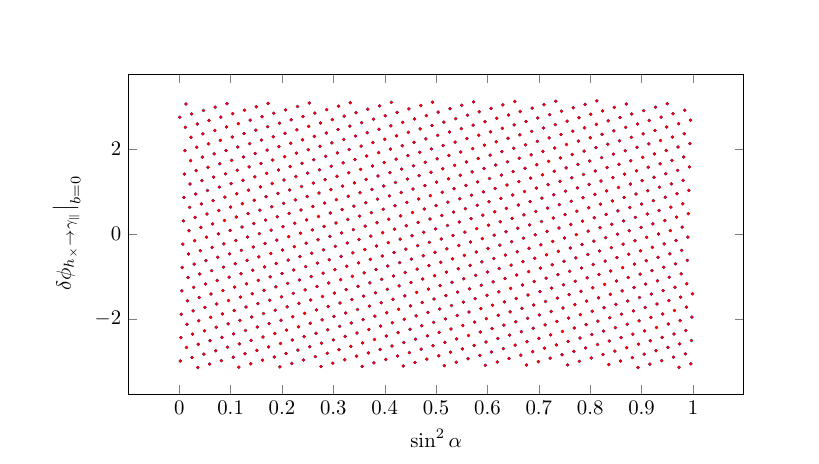}
    \end{subfigure}
    \hspace{-20pt}
    \begin{subfigure}{0.515\textwidth}
        \centering
        \includegraphics[width=\linewidth]{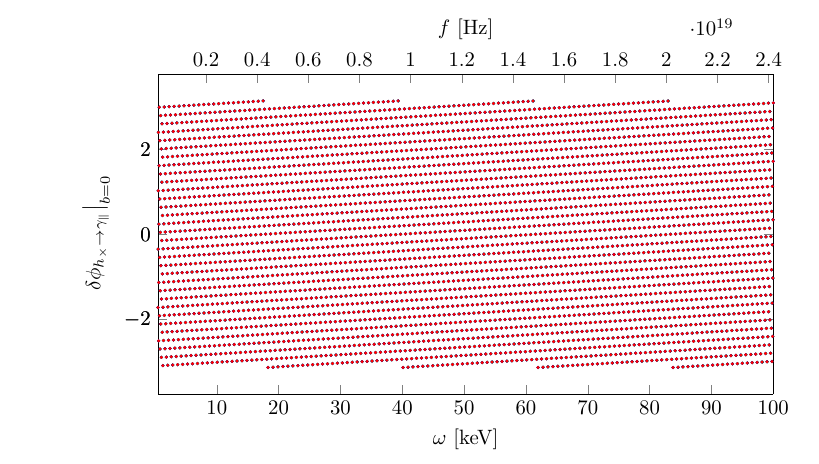}
    \end{subfigure}\\
    \begin{subfigure}{0.515\textwidth}
        \centering
        \includegraphics[width=\linewidth]{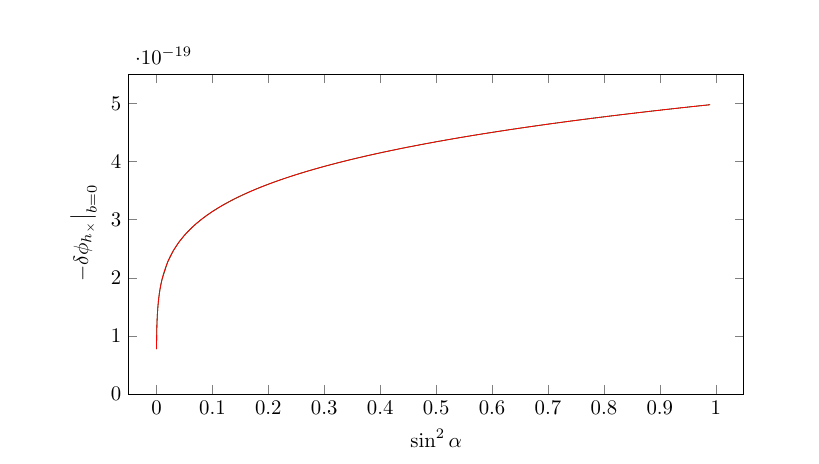}
    \end{subfigure}
    \hspace{-20pt}
    \begin{subfigure}{0.515\textwidth}
        \centering
        \includegraphics[width=\linewidth]{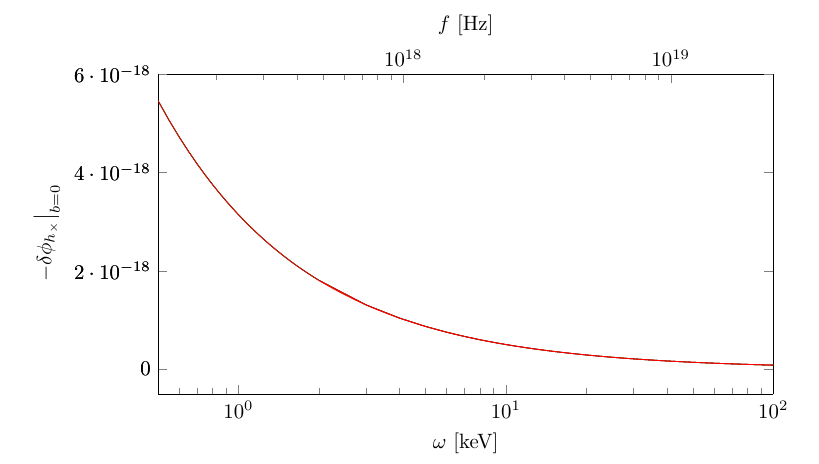}
    \end{subfigure}
    \caption{Conversion probability and phase shifts induced by the Gertsenshtein effect as functions of the polar angle relative to the magnetic moment $\alpha$ and angular frequency $\omega$. Green curves are obtained from the numerical integration of \eqref{NumericalProbInt} and \eqref{PhotonDiagPhaseshiftExactRadial}, blue curves from the exact expressions \eqref{MixingProbShiftRadialExact}, and red curves from the analytical approximations \eqref{ApproxProbShiftRadial} and \eqref{RadialApproxPhaseShift}. The three results nearly coincide, leaving only the red curves visible. The off-diagonal phase shifts oscillate rapidly due to the large value of $\lambda$, with those of \eqref{ApproxProbShiftRadial} defined modulo $2\pi$ in the interval $(-\pi,\pi)$. Benchmark parameters correspond to the magnetar 1E~1547.0$-$5408, with $r_0=10$ km and $B_0=3.2\times10^{14}\,\text{G}$. The left panels use $\omega=10\,\text{keV}$, while the right panels use $\alpha=\pi/2$.}
    \label{fig:RadialCaseApproxVsNumerical}
\end{figure}
We kept a higher precision in \eqref{RadialApproxPhaseShift} than in \eqref{ApproxProbShiftRadial} since the diagonal phase shift doesn't have an analytical closed form like \eqref{MixingProbShiftRadialExact}. The approximations \eqref{ApproxProbShiftRadial} and \eqref{RadialApproxPhaseShift} will be valid as long as $\lambda \gg 1$, which holds as long as $\sin \alpha$ is not so small as to compensate for the large value of $\Delta_{j0} r_0$. This occurs only for angles extremely close to $0$ or $\pi$, but these cases are of no interest, since the conversion probabilities and the diagonal phase shifts are almost zero due to the overall positive factor of $\sin \alpha$ in \eqref{MixingProbShiftRadialExact} and \eqref{PhotonDiagPhaseshiftExactRadial}.

Thus, \eqref{ApproxProbShiftRadial} and \eqref{RadialApproxPhaseShift} are reliable in the physically interesting region, as can be seen from Figure~\ref{fig:RadialCaseApproxVsNumerical} which shows the conversion probability and the phase shifts computed numerically, analytically with the exact solutions, and in the large $\lambda$ approximation
for the $\qty{A_\parallel, h_\times}$ system.


\subsection{Parallel Case} \label{sc:Parallel_Case}

For our second case of interest, we consider a distant GW source located at $z_0 \to -\infty$, whose radiation passes by the neutron star with impact parameter $b$ and is subsequently observed at $z \to \infty$. We focus on the specific case of an incident wave aligned with the magnetic moment, corresponding to $\alpha = 0$ in \eqref{DipoleMagneticField}. In this configuration, the magnetic field simplifies to
\begin{align}
    \left.\mathbf{B}\right|_{\alpha = 0} &= \frac{3B_0}{2} \frac{br_0^3z}{(z^2+b^2)^{\frac{5}{2}}}\qty(\cos\varphi, \sin\varphi, \frac{2z^2 - b^2}{3 b z}). \label{parallel_Bfield}
\end{align}
Here, $\tan\Phi =\frac*{B_y}{B_x} = \tan\varphi$ is constant. However, for negative $z$, we see that $\Phi$ is exactly opposite to $\varphi$ while it is equal to $\varphi$ for positive $z$. Hence, $\Phi$ changes quadrant when crossing $z = 0$, but this singular behavior can be cured by setting $\Phi = \varphi - \pi$ for $z<0$ and $\Phi = \varphi$ for $z>0$ in \eqref{EOM_perpar} and then join both equations of motion. This is equivalent to removing the absolute value from the transverse magnetic field $B_t = \sqrt{B_x^2 + B_y^2}$ in $\Delta_M$. Then, from \eqref{parallel_Bfield}, the transverse magnetic field $B_t$ can be computed and the $\Delta$'s appearing in the Hamiltonians \eqref{Hamiltonians} become
\begin{align}
    \left.\Delta_j\right|_{\alpha = 0} &=  \frac{9\Delta_{j0} b^2r_0^6 z^2}{(z^2 + b^2)^{5}}, & \left. \Delta_M\right|_{\alpha = 0} &=  \frac{3\Delta_{M0}br_0^3 z}{(z^2+b^2)^{\frac*{5}{2}}}, \label{ParallelDelta_Definition}
\end{align}
where $\Delta_{j0}$ and $\Delta_{M0}$ are the same as those found in \eqref{RadialDelta_Definition}.

\subsubsection{Conversion Probabilities and Phase Shifts}

As before, the conversion probabilities and the off-diagonal phase shifts are related to the off-diagonal element of the interaction evolution operator through \eqref{definition_probability_phaseshift}, given by
\begin{align}
    \left. \qty[\mathcal{U}_{\text{int}}]_{21}\right|_{\alpha = 0}&= -3i(\Delta_{M0}r_0)br_0^2   \int_{z_0}^z \dd{z'} \frac{z'}{({z'}^2 + b^2)^{\frac*{5}{2}}} \exp\qty[-9i\Delta_{j 0 }b^2r_0^6  \int_{z_0}^{z'} \dd{z''} \frac{{z''}^2}{({z''}^2 + b^2)^{5}}].
\end{align}
We change variables using $z' =b \tan \vartheta'$ and perform the same substitution for $z''$. Taking the source and the observer to be located at $z_0\to -\infty$ and $z\to \infty$, respectively, the resulting integral can be simplified using symmetry arguments, yielding the following expressions for the conversion probabilities and the off-diagonal phase shifts
\eqn{
\begin{gathered}
    \left.P_{h_j \to \gamma_j}\right|_{\alpha=0} = \frac{(6\Delta_{M0}r_0)^2}{{\bar b}^4}  \left|\int_0^{\frac{\pi}{2}} \dd{\vartheta'} \sin\vartheta' \cos^2\vartheta' \sin\qty(\frac{9\Delta_{j0 }r_0}{{\bar b}^5}\int_{0}^{\vartheta'} \dd{\vartheta''}\sin^2\vartheta'' \cos^6\vartheta''   ) \right|^2, \\
    \left.\delta \phi_{h_j\to \gamma_j}\right|_{\alpha=0} = -\frac{45 \pi \Delta_{j 0}r_0}{256 {\bar b}^5},
\end{gathered}
}[MixingProbShiftParallelExact]
where we introduced the adimensional impact parameter $\bar b = \frac*{b}{r_0}$. The diagonal phase shifts $\delta \phi_{h_j}$ is obtained by substituting \eqref{ParallelDelta_Definition} in \eqref{PerturbationAnalyticalApproximation}
\begingroup\makeatletter\def\f@size{10}\check@mathfonts\def\maketag@@@#1{\hbox{\m@th\large\normalfont#1}}
\eqn{
\begin{aligned}
    \left. \delta \phi_{h_j}\right|_{\alpha = 0} &= \\
    &\hspace{-25pt}(3\Delta_{M0}br_0^3 )^2\Im\left\{\int_{z_0}^z \dd{z'} \int_{z_0}^{z'}  \dd{z''}\frac{z'}{({z'}^2 + b^2)^{\frac*{5}{2}}} \frac{z''}{({z''}^2 + b^2)^{\frac*{5}{2}}} \exp\qty[-9i\Delta_{j 0}b^2r_0^6\int_{z''}^{z'}\dd{z'''} \frac{ {z'''}^2}{({z'''}^2 + b^2)^{5}}]\right\}.
\end{aligned}
}[PhaseshiftStartParallel]
\endgroup
This integral can also be simplified by applying the change of variables $z' = b\tan \vartheta'$ and performing the same substitution for $z''$ and $z'''$. Taking again $z_0\to -\infty$ and $z\to \infty$ for the source and the observer leads to
\begingroup\makeatletter\def\f@size{10}\check@mathfonts\def\maketag@@@#1{\hbox{\m@th\large\normalfont#1}}%
\eqn{
\begin{aligned}
     \left. \delta \phi_{h_j}\right|_{\alpha = 0} &= \\
     &\hspace{-25pt}-\frac{(3\Delta_{M0}r_0)^2}{{\bar b}^4} \int_{-\frac{\pi}{2}}^{\frac{\pi}{2}} \dd{\vartheta'}\int_{-\frac{\pi}{2}}^{\vartheta'}  \dd{\vartheta''} \sin\vartheta'\cos^2\vartheta' \sin\vartheta''\cos^2\vartheta'' \sin\qty(\frac{9\Delta_{j 0}r_0}{{\bar b}^5} \int_{\vartheta''}^{\vartheta'} \dd{\vartheta'''}\sin^2 \vartheta''' \cos^6\vartheta''').
\end{aligned}
}[PhaseshiftExactParallel]
\endgroup
As in the radial case, the phase shift induced by nonlinear QED effects, $\delta \phi_{\text{QED}}$, follows directly from \eqref{deltaphi_QED} and \eqref{ParallelDelta_Definition}, yielding
\begin{align}
    \left.\delta \phi_{\text{QED}}\right|_{\alpha=0} &= \frac{45\pi}{128 \bar b^5}\qty(\Delta_{\perp 0}r_0- \Delta_{\parallel 0 }r_0)=  -\frac{3 \alpha \omega r_0 }{1024 \bar b^5}\qty(\frac{B_0}{B_{\text{crit}}})^2.  \label{PhaseShiftQEDParallel}
\end{align}

\subsubsection{Analytical Approximations}

\begin{figure}[t!]
    \begin{subfigure}{0.505\textwidth}
        \centering
        \includegraphics[width=\linewidth]{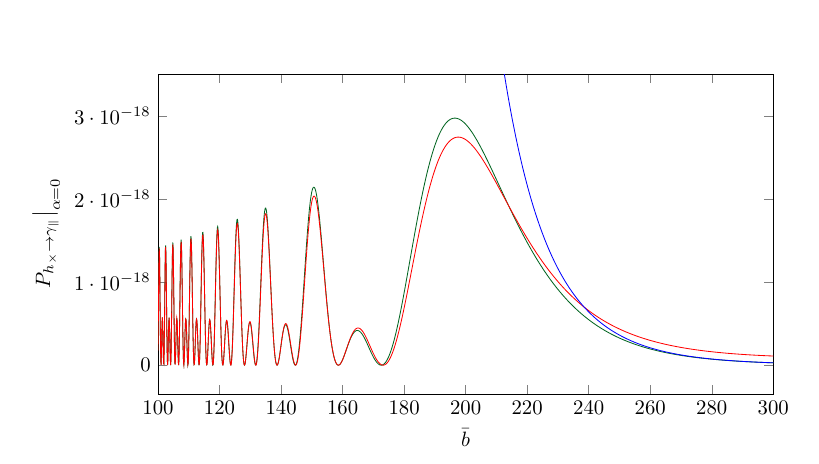}
    \end{subfigure}
    \hspace*{-12.5pt}
    \begin{subfigure}{0.505\textwidth}
        \centering
        \includegraphics[width=\linewidth]{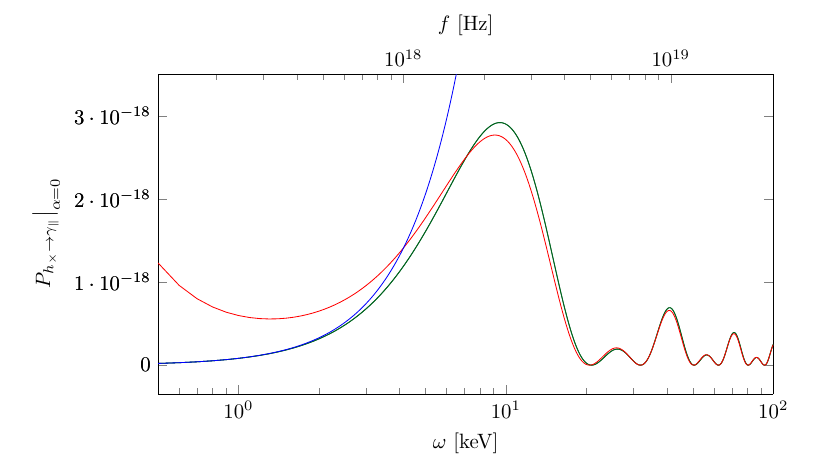}
    \end{subfigure} \\
    \begin{subfigure}{0.505\textwidth}
        \centering
        \includegraphics[width=\linewidth]{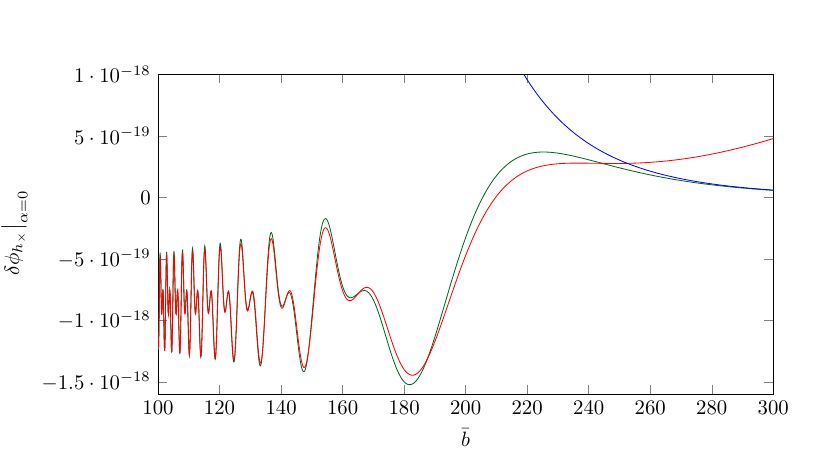}
    \end{subfigure}
    \hspace*{-12.5pt}
    \begin{subfigure}{0.505\textwidth}
        \centering
        \includegraphics[width=\linewidth]{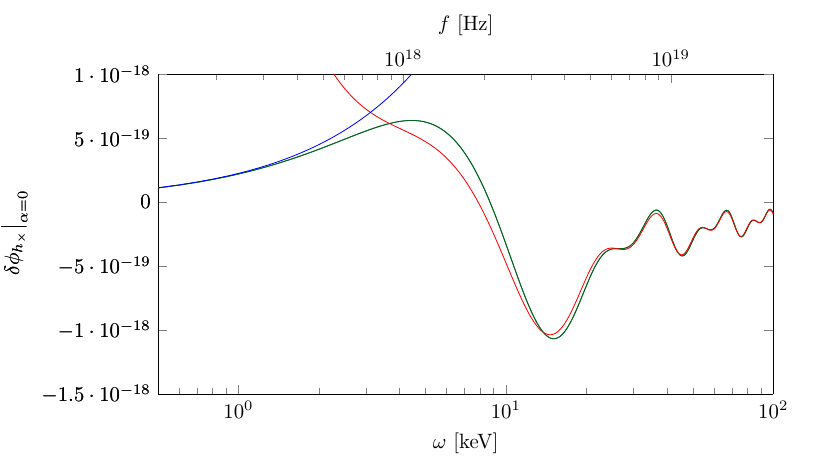}
    \end{subfigure}
    \caption{Conversion probability and diagonal phase shift induced by the Gertsenshtein effect as functions of the dimensionless impact parameter $\bar b = \frac*{b}{r_0}$ and angular frequency $\omega$. The green curves are obtained from numerical integration of \eqref{MixingProbShiftParallelExact} and \eqref{PhaseshiftExactParallel}, the red curves correspond to the analytical approximations \eqref{ParallelApproxbsimr0} and \eqref{deltaphihjParallelApproxbsimr0} in the regime $b\sim r_0$, and the blue curves denote the asymptotic approximations \eqref{LargebParallelApprox} valid for $b\gg r_0$, where the conversion probability rapidly decreases. Benchmark magnetar parameters are $r_0 = 10$ km and $B_0 = 3.2\times 10^{14}\, \text{G}$, corresponding to 1E~1547.0-5408. The left panels use $\omega = 10\, \text{keV}$, while the right panels use $\bar b = 200$.}
    \label{fig:ParallelCaseApproxVsNumerical}
\end{figure}

The conversion probabilities and diagonal phase shifts in \eqref{MixingProbShiftParallelExact} and \eqref{PhaseshiftExactParallel} are not yet in analytical form, which motivates the development of approximations in two different regimes. The first natural regime occurs when the argument of the sine functions is much smaller than one, which happens when the impact parameter is much larger than the neutron star radius ($b\gg r_0$). In this limit, the sine functions in the integrals can be approximated by the first term of their Taylor series, allowing the integrals to be evaluated explicitly and yielding
\eqn{
\begin{gathered}
    \left.P_{h_j \to \gamma_j}\right|_{\scalebox{0.67}{$\mqty{\hspace{-4pt}\alpha = 0 \vspace{-5pt}\\ \hspace{1pt} b \gg r_0}$}}
    \approx  \qty[\frac{256 (\Delta_{M0}r_0) (\Delta_{j 0 }r_0)}{385 {\bar b}^7}]^2 , \\
      \left. \delta \phi_{h_j}\right|_{\scalebox{0.67}{$\mqty{\hspace{-4pt}\alpha = 0 \vspace{-5pt}\\ \hspace{1pt} b \gg r_0}$}} \approx \frac{11\pi \Delta_{j 0}r_0}{6 {\bar b}^9}\qty(\frac{9\Delta_{M0}r_0}{32})^2.
\end{gathered}
}[LargebParallelApprox]
This asymptotic behavior is unique to the parallel case since a series expansion in any other orientation will yield conversion probabilities and diagonal phase shifts that all decay as ${\bar b}^{-4}$ as already pointed out in \cite{Dandoy:2024oqg}. However, since the contribution from $\Delta_\Phi r_0$ in \eqref{EOM_perpar} is dominant in other orientations, it should be retained for a rigorous treatment. Nonetheless, the conclusion remains the same, the conversion probabilities and the diagonal phase shifts are strongly suppressed when the incident ray passes far from the neutron star. Consequently, the interesting region  where conversion may occur at an observable level corresponds to the opposite regime where the argument of the sine functions is much larger than one, which happens for $b\sim r_0$. To obtain an approximative solution in this regime, we again exploit the fact that $\Delta_{j 0}r_0$ is a large dimensionless parameter for typical neutron stars, as explained at the beginning of Subsection~\ref{AnalyticalApproximationsRadial}. Hence for moderate value of the impact parameter $b \in \qty[r_0, 100r_0]$ the large-argument criterion is met and one can use the stationary phase method for degenerate critical points to approximate the conversion probabilities \eqref{MixingProbShiftParallelExact} and the diagonal phase shifts \eqref{PhaseshiftExactParallel}. Thus, using the method presented in Appendix~\ref{Appendix:SPMethod} with
\begin{align}
    f(\vartheta) &=  \frac{256}{5\pi}\int^{\vartheta}_{0} \dd{\vartheta'}\sin^2 \vartheta' \cos^6\vartheta', & g(\vartheta) &= \sin\vartheta \cos^2\vartheta, & \lambda &= \frac{45\pi \Delta_{j0}r_0}{256{\bar b}^5}, \label{fgthetalambdaDefinition}
\end{align}
one can separate the integration domain appropriately and sum over the contributions of the critical points $\vartheta_* = \qty{-\frac*{\pi}{2}, 0, \frac*{\pi}{2}}$ of $f(\vartheta)$ in the integration domain of \eqref{MixingProbShiftParallelExact} and \eqref{PhaseshiftExactParallel} to obtain
\begingroup\makeatletter\def\f@size{10}\check@mathfonts\def\maketag@@@#1{\hbox{\m@th\large\normalfont#1}}%
\begin{align}
         \left.P_{h_j \to \gamma_j}\right|_{\scalebox{0.67}{$\mqty{\hspace{-4pt}\alpha = 0 \vspace{-5pt}\\ \hspace{1pt} b \sim r_0}$}} &\approx  \nonumber  \\
        &\hspace{-35pt} \frac{\qty(\Delta_{M0}r_0)^2}{{\bar b}^4} \qty|\frac{3\qty(\frac*{35\pi}{2})^{\frac{3}{7}}\Gamma\qty(\frac{3}{7})\cos\qty(\lambda - \frac{5\pi}{7})}{28\lambda^{\frac{3}{7}}}  + \frac{3^{\frac{1}{2}} \qty(30\pi)^{\frac{2}{3}}\Gamma\qty(\frac{2}{3}) }{64\lambda^{\frac{2}{3}}}-\frac{5\qty(\frac*{35\pi}{2})^{\frac{5}{7}}\Gamma\qty(\frac{5}{7})\cos\qty(\lambda + \frac{\pi}{7})}{672\lambda^{\frac{5}{7}}} + \mathcal{O}\qty[\lambda^{-1}]|^2,   \label{ParallelApproxbsimr0}
\end{align}
\vspace{-25pt}
\begin{align}
    \left. \delta \phi_{h_j}\right|_{\scalebox{0.67}{$\mqty{\hspace{-4pt}\alpha = 0 \vspace{-5pt}\\ \hspace{1pt} b \sim r_0}$}} &\approx \nonumber \\
    &\hspace{-35pt}-\frac{(3\Delta_{M0}r_0)^2 }{{\bar b}^4} \left\{\frac{\qty(\frac*{35\pi}{2})^{\frac{6}{7}}\Gamma\qty(\frac{3}{7})^2\qty[\tan\qty(\frac{3\pi}{7})-\sin\qty(2\lambda - \frac{3\pi}{7})]}{3136\lambda^{\frac{6}{7}}}    - \frac{\qty(\frac*{35\pi}{2})^{\frac{3}{7}}\qty(30\pi)^{\frac{2}{3}}\Gamma\qty(\frac{3}{7})\Gamma\qty(\frac{2}{3})\sin\qty(\lambda + \frac{2\pi}{7})}{1792\cdot 3^{\frac{1}{2}}\lambda^{\frac{23}{21}} } \right. \nonumber \\
    &\hspace{33pt}\left.- \frac{5\qty(\frac*{35\pi}{2})^{\frac{8}{7}}\Gamma\qty(\frac{3}{7})\Gamma\qty(\frac{5}{7})\qty[\cos\qty(\frac{\pi}{7})\tan\qty(\frac{3\pi}{7}) + \sin\qty(2\lambda-\frac{4\pi}{7} )]}{112\,896\lambda^{\frac{8}{7}}} - \frac{\qty(30\pi)^{\frac{4}{3}}\Gamma\qty(\frac{2}{3})^2}{8192\cdot 3^{\frac{1}{2}}\lambda^{\frac{4}{3}} } + \mathcal{O}\qty[\lambda^{-\frac{29}{21}}]\right\}.   \label{deltaphihjParallelApproxbsimr0}
\end{align}
\endgroup
In the computation of \eqref{ParallelApproxbsimr0} and~\ref{deltaphihjParallelApproxbsimr0}, we took the choice to truncate terms of order beyond $\lambda^{-\frac*{4}{3}}$. This is justified by the fact that we wanted to keep the first-order contribution of each critical point. A comparison between the values obtained from numerically integrating \eqref{MixingProbShiftParallelExact} and~\ref{PhaseshiftExactParallel} and the analytical approximations presented at \eqref{ParallelApproxbsimr0} and \eqref{deltaphihjParallelApproxbsimr0} is shown in Figure~\ref{fig:ParallelCaseApproxVsNumerical}.

From Figure~\ref{fig:ParallelCaseApproxVsNumerical}, two characteristic values of the impact parameter are of particular interest. The first corresponds to the maximal conversion, which can be estimated from the peak of the final oscillation in the probability. This is well approximated by setting the argument of the cosine in the dominant term of \eqref{ParallelApproxbsimr0} to zero, yielding
\begin{align}
     b_{\text{max}} \approx \qty(\frac{63\Delta_{j0}r_0}{256})^{\frac{1}{5}}r_0. \label{maximal_impact_parameter}
\end{align}
The second characterizes the transition between the regimes $b \sim r_0$ and $b \gg r_0$, corresponding to $\lambda = 1$. Since the conversion probabilities are strongly suppressed outside the region $b \sim r_0$, this transition effectively marks the onset of efficient conversion. It therefore defines the radius of conversion, providing an analytical estimate of the critical impact parameter $b_c$ introduced in \cite{Dandoy:2024oqg} for this peculiar geometry:
\begin{align}
    b_c &= \qty(\frac{45\pi \Delta_{j0}r_0}{256})^{\frac{1}{5}}r_0. \label{critical_impact_parameter}
\end{align}
For typical magnetar and X-ray parameters, both impact parameters lie well outside the neutron star surface. Using the benchmark values of 1E~1547.0-5408 given in Figure~\ref{fig:ParallelCaseApproxVsNumerical}, we find $b_{\text{max}} \approx 197r_0$ and $b_c \approx 231r_0$, indicating that the conversion predominantly occurs far from the magnetar's surface where the dipolar approximation is justified. The same reasoning can also be applied to the diagonal phase shift $\delta \phi_{h_j}$ given in \eqref{deltaphihjParallelApproxbsimr0}, leading to similar physical implications.


\section{Bounds on the Gravitational Wave Background}\label{SecBounds}

In this section, we will compute lower and upper bounds on the SGWB using the magnetospheres of magnetars as regions of conversion.  The lower bounds will come from Gertsenshtein conversion of photons emitted by magnetars while the upper bounds will be obtained from inverse Gertsenshtein conversion of the SGWB to photons \cite{Ito:2023fcr, Dandoy:2024oqg}.

As discussed previously, we will only consider neutron stars whose magnetic moment is aligned with their rotation axis. Although this assumption is strongly disfavored by the existence of pulsars, it remains a standard approximation for analytical treatments. In a more realistic configuration, however, where the magnetic field is not perfectly aligned with the rotation axis, the azimuthal angle $\Phi$ acquires a $z$-dependence, even for wave packets propagating radially to the magnetic moment. As a result, $\Delta_\Phi \neq 0$, and given the smallness of $\Delta_M$, $\Delta_\Phi$ is expected to dominate even for very small misalignments as mentioned in Subsection \ref{ssc:EOM}.

This observation suggests that a precise treatment of the system likely requires a numerical analysis in $x$-$y$ coordinates, as done by \cite{ejlli_graviton-photon_2019,  Chiba:2025odu, Gupta:2026scx}. Nevertheless, the term $\Delta_\Phi$ in \eqref{EOM_perpar} only mixes the photon polarization states and the gravitational-wave polarization states among themselves, while photon--gravitational-wave conversion remains mediated solely by the Gertsenshtein coupling $\Delta_M$. Consequently, the conversion probabilities \eqref{PerturbationAnalyticalApproximation} are expected to remain unchanged to leading order in $\Delta_M$ and $\Delta_\Phi$, as was shown for the axion case by \cite{Wang:2015dil, Seong:2023ran}. Since the Stokes parameters $I_\gamma$ and $I_h$ defined in \eqref{photonStokesParameter} and \eqref{gravitonStokesParameter} are insensitive to polarization mixing, their modifications are expected to arise primarily from the magnetic-field orientation along a generic trajectory. Because the magnetosphere is modelled by a dipolar magnetic field with similar magnitude and radial scaling in all directions, we expect the resulting conversion probabilities to remain of the same order of magnitude even beyond the adiabatic approximation.

In light of this, we adopt simplified assumptions and examine the two geometries for which the mixing term $\Delta_\Phi = 0$, to derive upper and lower bounds using the results of the previous sections.


\subsection{Lower Bounds}

Neutron stars emit X-rays and host strong magnetic fields, and are therefore expected to produce HFGWs. In this section, we estimate the order of magnitude of this emission, even though it is anticipated to be extremely small. This provides a physically motivated prediction for gravitational waves generated via Gertsenshtein conversion of radially escaping X-rays in the magnetosphere, and thus a lower bound on the characteristic strain in this frequency range.

\begin{figure}[t!]
    \centering
    \includegraphics[width=\linewidth]{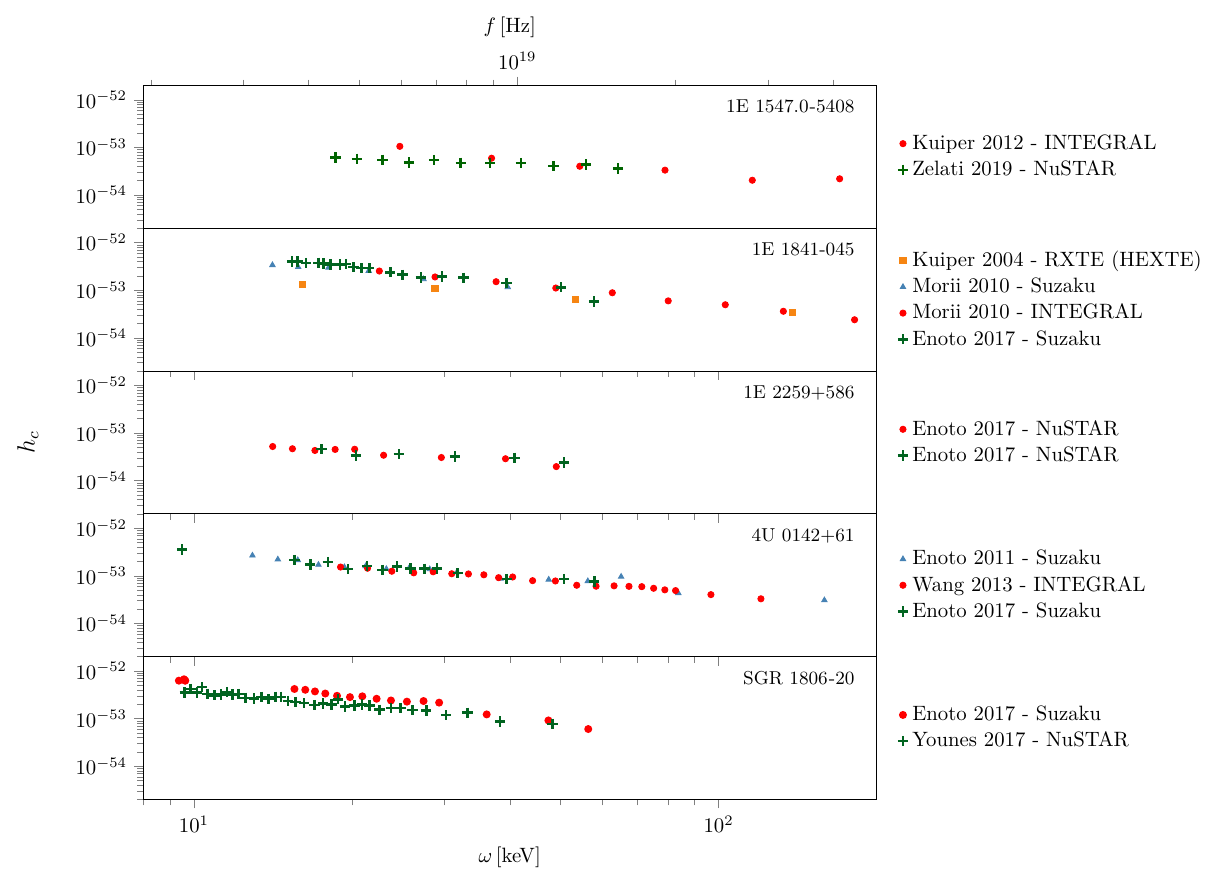}
    \caption{Characteristic strain measured at Earth from gravitational waves produced through Gertsenshtein conversion of radially propagating X-rays in magnetars' magnetospheres.}
    \label{fig:LowerBounds}
\end{figure}

We assume the emitted radiation is unpolarized, $Q_{\gamma 0 } = U_{\gamma 0 } = V_{\gamma 0 }=0$, and that no other sources contribute to HFGWs, such that all initial gravitational-wave Stokes parameters vanish. Under these assumptions, the gravitational-wave Stokes parameter $I_h$, given in \eqref{gravitonStokesParameter}, reduces to
\eqn{
\begin{aligned}
   I_{h} &= \frac{I_{\gamma 0}}{2} \sum_j \left[ P_{h_j \to \gamma_j}\right|_{b=0} .
\end{aligned}
}[RadialStokesParameters]
However, the Stokes parameters above are defined in the perpendicular--parallel basis with respect to the transverse magnetic field, \textit{i.e.}\ in terms of $A_\perp, A_\parallel, h_+$ and $h_\times$. This basis depends on the azimuthal angle of the magnetic field, $\Phi= \beta-\pi$, as shown in \eqref{PerpParBasis} and \eqref{radial_Bfield}, and therefore varies between sources. Furthermore, the magnetic moment orientation in reference to Earth, parameterized by the angles $\alpha$ and $\beta$, is not always known and difficult to constrain observationally \cite{NgRomani:2008}. We therefore introduce a fixed $x$-$y$ reference frame on Earth and average over the magnetic moment inclination compared to Earth, treating it as effectively random. The relevant Stokes parameter is then
\begin{align}
     I_{h}^{(\text{$x$-$y$})} &= \frac{1}{4\pi}\int \dd{\Omega_{\mathbf{\hat{m}}}} \int_0^{2\pi} \frac{\dd{\Delta \phi_{\gamma_{\perp0}}}}{2\pi} \qty[ \qty|h_{xx}|^2 + \qty|h_{xy}|^2 ], \nonumber  \\
     &= I_{\gamma 0}\sum_{j}\frac{1}{8\pi}\int_0^{2\pi}\dd{\beta} \int_{-1}^{1} \dd{\cos\alpha}  \left[ P_{h_j \to \gamma_j}\right|_{b=0}=I_{\gamma 0}\bar P_{h \to \gamma}^{(\text{rad})}, \label{StokesRadialRelationGraviton}
\end{align}
where the last line follows from expressing $h_{xx}$ and $h_{xy}$ in terms of $\Phi= \beta-\pi$ [see \eqref{radial_Bfield}] using the change of bases \eqref{PerpParBasis}. The total averaged radial conversion probability, which multiplies the initial photon Stokes parameter $I_{\gamma 0}$ in \eqref{StokesRadialRelationGraviton}, is computed using the approximation \eqref{ApproxProbShiftRadial}
\begin{align}
    \bar P_{h \to \gamma}^{(\text{rad})} &\approx \sum_j\frac{\Gamma\qty(\frac{2}{5})\Gamma\qty(\frac{1}{5})^2\qty(\Delta_{M0}r_0)^2}{70 \cdot 2^{\frac{3}{5}}\cdot 5^{\frac{1}{5}} \qty(\Delta_{j 0 } r_0)^{\frac{4}{5}}}, \nonumber \\
    &\approx 1.7\times10^{-19}\cdot \qty(\frac{B_0}{B_{\text{crit}}})^{\frac{2}{5}} \qty(\frac{r_0}{10\, \text{km}})^{\frac{6}{5}}\qty(\frac{10\, \text{keV}}{\omega})^{\frac{4}{5}}. \label{AverageRadialConversionProbability}
\end{align}
The final expression follows from substituting the $\Delta$’s with their explicit forms given in \eqref{RadialDelta_Definition}. From the stress-energy tensor, the Stokes parameters $I$ are seen to be proportional to the radial energy flux of each field. Accordingly, \eqref{StokesRadialRelationGraviton} may be equivalently expressed in terms of the differential energy flux, following \cite{Ito:2023fcr, ramazanov_shimmering_2023, Dandoy:2024oqg}:
\begin{align}
    \pdv{F_{h}}{\omega} &=  \bar P_{h \to \gamma}^{(\text{rad})}\pdv{F_{\gamma 0}}{\omega}. \label{FluxRelationRadial}
\end{align}
In practice, the photon energy flux from a neutron star is typically expressed in terms of $\nu F_\nu$, which is related to the differential photon energy flux by \cite{Fortin:2021sst}
\begin{align}
    \nu F_\nu (\omega) &= \omega \pdv{F_{\gamma 0}}{\omega}. \label{nuFnuDefinition}
\end{align}
The gravitational-wave energy flux is usually expressed in terms of the characteristic strain $h_c$, which is related to the differential energy flux (for radially propagating waves in all directions) through \cite{Maggiore:1999vm, Dandoy:2024oqg}
\begin{align}
    \pdv{F_h}{\omega}  &= \frac{m_P^2}{2}\omega h_c^2. \label{GravitationalWaveFluxStrain}
\end{align}
Combining \eqref{AverageRadialConversionProbability}, \eqref{nuFnuDefinition}, and \eqref{GravitationalWaveFluxStrain} into \eqref{FluxRelationRadial}, we obtain the expected characteristic strain measured at Earth from the Gertsenshtein conversion of X-rays in a magnetar's magnetosphere
\begin{align}
    h_c &\gtrsim 3.8\times 10^{-53}\cdot \qty(\frac{B_0}{B_{\text{crit}}})^{\frac{1}{5}}\qty(\frac{\omega}{10\,\text{keV}})^{-\frac{7}{5}}  \qty(\frac{r_0}{10\,\text{km}})^{\frac{3}{5}}\qty(\frac{\nu F_\nu}{10^{-2}\frac{\text{keV}^2\cdot \text{Photons} }{\text{cm}^2 \cdot \text{s} \cdot \text{keV}}})^{\frac{1}{2}}. \label{RadiatedStrainNeutronStar}
\end{align}
Using \eqref{RadiatedStrainNeutronStar} together with experimental measurements of the $\nu F_\nu$ X-ray spectra of magnetars, we derive lower bounds on the characteristic strain of HFGWs. To this end, we use the spectral data of five magnetars, namely 1E 1547.0-5408, 1E 1841-045, 1E 2259+586, 4U 0142+61, and SGR 1806-20, to estimate the gravitational-wave signal produced through Gertsenshtein conversion of X-rays (see Figure~\ref{fig:LowerBounds}). Additional details regarding the magnetar spectra and parameters used here are provided in Appendix~\ref{Appendix:MagnetarsParameters}.

\subsection{Upper Bounds}

To obtain upper bounds, we consider a HFGW from the background passing near a neutron star, where it would be partially converted into photons via the inverse Gertsenshtein effect.

Assuming an unpolarized gravitational wave background, we can set the initial Stokes parameters of the gravitational wave $Q_{h0}=U_{h0} = V_{h0} = 0$. For the photon, we assume that no significant background sources are present, so that all initial Stokes parameters are zero which should be reasonable over the frequency range of X-rays. In this specific case, the photon Stokes parameter $I_\gamma$ is given by \eqref{photonStokesParameter}:
\begin{align*}
     I_{\gamma} &= \frac{I_{h 0}}{2} \sum_j P_{h_j \to \gamma_j}.
\end{align*}
As noted above, the orientation of the magnetic moment of a neutron star with respect to Earth is difficult to determine observationally \cite{NgRomani:2008}. To simplify the analysis, we assume that the order of magnitude of the conversion for arbitrary inclinations is comparable to that of the parallel configuration ($\alpha=0$), following the discussion in the introduction to this section.

\begin{figure}[t!]
    \centering
    \includegraphics[width=\linewidth]{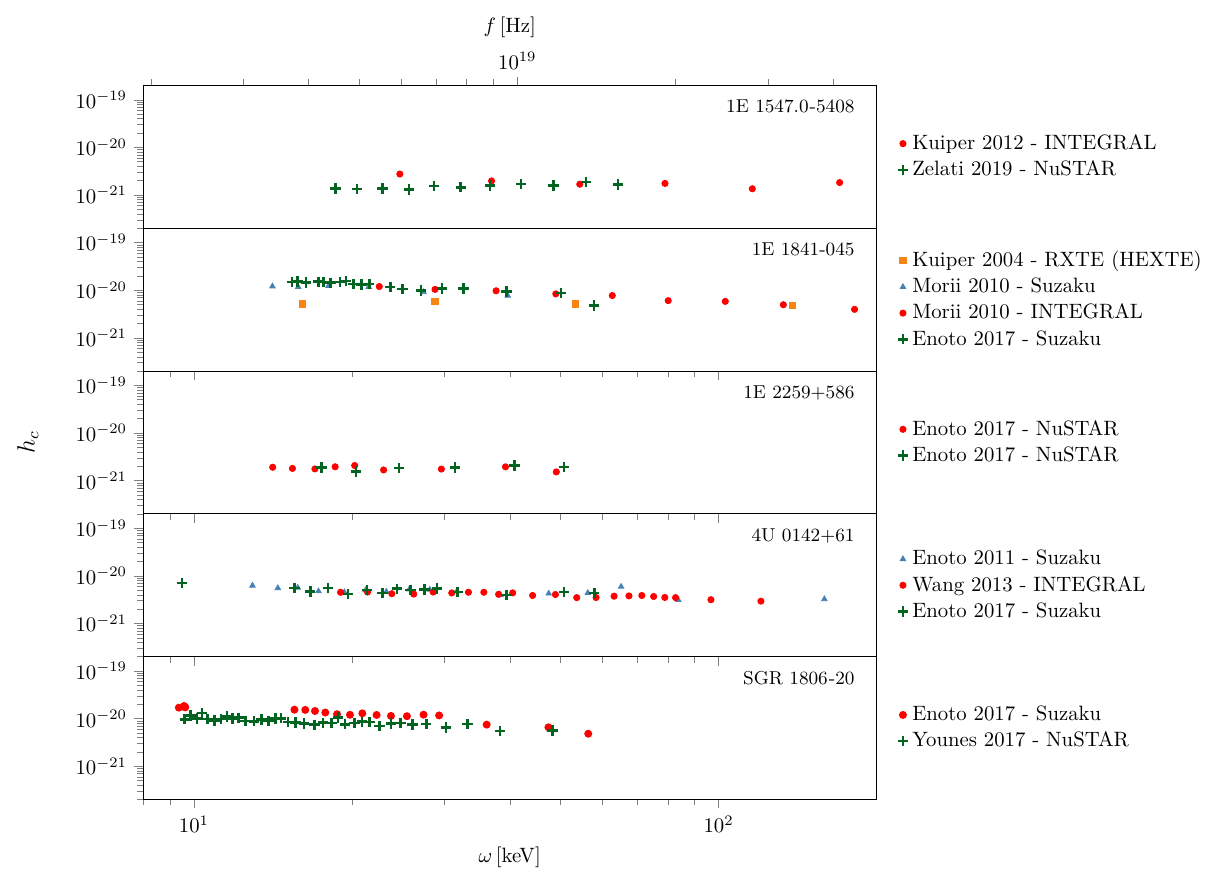}
    \caption{Upper bounds on the characteristic strain of the gravitational-wave background derived from Gertsenshtein conversion in magnetars' magnetospheres. }
    \label{fig:UpperBounds}
\end{figure}

As discussed previously, the Stokes parameter $I_\gamma$ is defined in the perpendicular--parallel basis relative to the transverse magnetic field. This basis is only uniquely specified for a given neutron star upon replacing $\Phi = \varphi$ in \eqref{PerpParBasis}. The Stokes parameters observed in the Earth-based $x$-$y$ frame are therefore given by
\begin{align*}
    I_\gamma^{(\text{$x$-$y$})} &=  \int_0^{2\pi} \frac{\dd{\Delta \phi_{\gamma_{\perp0} h_{\times 0}}}}{2\pi} \qty[ \qty|A_x(z)|^2 +  \qty|A_y(z)|^2  ], \\
    &=  \frac{I_{h 0}}{2} \sum_j \left[P_{h_j \to \gamma_j}\right|_{\alpha = 0},
\end{align*}
where the last line follows from expressing $A_{x}$ and $A_{y}$ with $\Phi= \varphi$ [see \eqref{parallel_Bfield}] using the transformation \eqref{PerpParBasis}. As before, the Stokes parameters $I$ are proportional to the energy flux in the $z$-direction, such that the above relation can equivalently be written as
\begin{align*}
    \pdv[2]{F_\gamma}{\Omega}{\omega} &= \frac{1}{2}\pdv[2]{F_{h0}}{\Omega}{\omega}  \sum_j\left[P_{h_j \to \gamma_j}\right|_{\alpha = 0}.
\end{align*}
Integrating both sides over all impact parameters $b$ yields the differential luminosity per unit solid angle of the emitted radiation. Assuming isotropy of both the conversion process and the background, the angular integration is trivial and gives the total differential luminosity associated with gravitational-wave background conversion. This result is then expressed in terms of the flux measured at Earth as follows,
\begin{align}
    \pdv{F_\gamma}{\omega} &= \frac{\sigma_{\text{eff}}}{4\pi d^2}\pdv{F_{h0}}{\omega}, \label{CrossSectionFluxRelation}
\end{align}
where $d$ is the distance to the neutron star and $\sigma_{\text{eff}}$ denotes the effective cross section of the conversion region:
\begin{align}
    \sigma_{\text{eff}} &= \frac{1}{2}\sum_j \int\dd[2]{b} \left[P_{h_j \to \gamma_j}\right|_{\alpha = 0}.
\end{align}
This quantity can be evaluated approximately analytically using the approximations \eqref{ParallelApproxbsimr0} and \eqref{LargebParallelApprox} in their respective domains. This process is detailed in Appendix \ref{Appendix:cte} and yields a power-law dependence over the parameters $\Delta_{j0}r_0$ and $\Delta_{M0}r_0$
\begin{align}
    \sigma_{\text{eff}} &\approx 1.05\cdot \sum_j \frac{\pi r_0^2 (\Delta_{M0}r_0)^2}{(\Delta_{j 0 }r_0)^{\frac{2}{5}}}, \label{ApproximateEffectiveCrossSectionDeltas}  \\
    &\approx 5.5\times 10^{13}\, \text{keV}^{-2} \cdot  \qty(\frac{B_0}{B_{\text{crit}}})^{\frac{6}{5}} \qty(\frac{\omega}{10\, \text{keV}})^{-\frac{2}{5}} \qty(\frac{r_0}{10\, \text{km}})^{\frac{18}{5}}, \label{ApproximateEffectiveCrossSection}
\end{align}
where the final expression follows from substituting the $\Delta$’s with their explicit forms given in \eqref{RadialDelta_Definition}. Combining the approximative effective cross section \eqref{ApproximateEffectiveCrossSection} with \eqref{CrossSectionFluxRelation}, and expressing the gravitational-wave background flux in terms of the characteristic strain via \eqref{GravitationalWaveFluxStrain} and the photon flux using \eqref{nuFnuDefinition}, one can constrain the characteristic strain by requiring that the flux observed at Earth generated through inverse Gertsenshtein conversion of the gravitational-wave background does not exceed the observed flux from the magnetar. This condition reads
\begin{align}
    h_c &\lesssim 3.5\times 10^{-20}\cdot\qty(\frac{B_0}{B_{\text{crit}}})^{-\frac{3}{5}} \qty(\frac{\omega}{10\,\text{keV}})^{-\frac{4}{5}} \qty(\frac{r_0}{10\,\text{km}})^{-\frac{9}{5}} \qty(\frac{d}{5\,\text{kpc}}) \qty(\frac{\nu F_\nu}{10^{-2}\,\frac{\text{keV}^2\cdot \text{Photons} }{\text{cm}^2 \cdot \text{s} \cdot \text{keV}}})^{\frac{1}{2}}.
    \label{FinalConstrainhcParallel}
\end{align}
Using \eqref{FinalConstrainhcParallel} together with the X-ray spectra of the magnetars 1E 1547.0-5408, 1E 1841-045, 1E 2259+586, 4U 0142+61, and SGR 1806-20, we derive the upper bounds on the characteristic strain of the gravitational-wave background shown in Figure~\ref{fig:UpperBounds}. Additional details regarding the magnetar spectra and specific parameters used in this analysis are provided in Appendix~\ref{Appendix:MagnetarsParameters}.


\section{Conclusion}\label{SecConc}

In this paper we took advantage of the Gertsenshtein effect to investigate the stochastic gravitational wave background in the X-ray frequency band with the help of magnetars. To this end we rederived the evolution equations of the photon--graviton system and pointed out that the latter do not generally decouple into two independent systems: the adiabatic approximation is not justified due to the smallness of the inverse Planck mass.  By investigating two specific geometries for which the adiabatic approximation is valid, we solved the evolution equations of the Stokes parameters in terms of the conversion probabilities and the phase shifts.  We then computed these quantities analytically for the two geometries of interest: radially emitted electromagnetic waves from the magnetar converted into gravitational waves; and background gravitational waves propagating parallel to the magnetic moment converted into electromagnetic waves in the magnetosphere.  With these results, we finally derived analytical lower and upper bounds, respectively, on the stochastic gravitational wave background in the X-ray frequency band.  Our lower bound demonstrates that the Gertsenshtein effect is not efficient in creating a stochastic gravitational wave background at those high frequencies, in agreement with the non-existence of standard cosmological events generating such a background.  Our upper bound, which is of the order of the observed characteristic strain at LIGO but in the $10^{18}-10^{20}\,\text{Hz}$ range, is obtained by constraining the Gertsenshtein-induced photon flux to not exceed the observed magnetar flux.

By including extra contributions, like plasma effects, it would be easy to improve our bounds to a wider range of frequencies.  This improvement may give competitive bounds on the characteristic strain in the radiowave frequency, where some magnetars are relatively luminous.  Moreover, considering that the axis of rotation of a magnetar is not necessarily aligned with its magnetic moment, it would be interesting to generalize our computations away from the adiabatic approximation regime.  Such a generalization would mix all four states (the perpendicular and parallel photon modes, and the plus and cross gravitational modes), leading to a much more complicated analysis.  We hope to extend our work to this more realistic scenario. Nevertheless, we believe that the overall analytical behaviors and orders of magnitude of the conversion probabilities and phase shifts described here should be correct.


\ack{
JSC is supported by the Natural Sciences and Engineering Research Council of Canada (NSERC) and the Fonds de recherche du Québec (FRQ) through awards \href{https://nserc-crsng.canada.ca/fr/base-donnees-subventions-bourses/797573}{596780-2024} and \href{https://doi.org/10.69777/371340}{371340}, respectively. The work of JFF is supported by NSERC.
}


\setcounter{section}{0}
\renewcommand{\thesection}{\Alph{section}}

\section{Stationary Phase Method for Degenerate Critical Points} \label{Appendix:SPMethod}

Here, we describe how to compute the contribution of critical points using the stationary phase method for degenerate critical points \cite{bleistein1986asymptotic}. We focus on two integral structures that frequently arise in the Dyson series of the form of
\eqref{SecondOrderUint}:
\begin{equation}
    \begin{gathered}
        \mathcal{I}_1\qty(\vartheta_*) = \int_{\vartheta_*}^{\delta} \dd{\vartheta'} g(\vartheta')e^{i \lambda f(\vartheta')}, \\
        \mathcal{I}_2\qty(\vartheta_*) = \int_{\vartheta_*}^{\delta} \dd{\vartheta'} \int_{\vartheta_*}^{\vartheta'} \dd{\vartheta''} g(\vartheta')g(\vartheta'')e^{i \lambda \qty[f(\vartheta')-f(\vartheta'')]}.
    \end{gathered} \label{IntSPMethod}
\end{equation}
where, w.l.o.g.\ we assume $\delta > \vartheta_*$ (otherwise, this can be enforced by a change of variables). We make the following assumptions:
\begin{enumerate}
    \item The functions $g(\vartheta')$ and $f(\vartheta')$ are real-valued and smooth on the integration domain, so that they admit power series expansions around $\vartheta_*$.
    \item The critical point $\vartheta_*$ is of finite order and is the unique critical point of $f(\vartheta')$ in the integration domain.
    \item The parameter $\lambda$ is sufficiently large so that rapid oscillations occur away from the neighborhood of the critical point, leading to strong cancellations in those regions. This is equivalent to the condition $\qty|\lambda \Delta f|\gg 1$, where $\Delta f = f(\delta) - f(\vartheta_*)$. In general, $\lambda$ should be chosen so that $\Delta f = 1$ which simplifies the  previous relation to $\qty|\lambda| \gg 1$.
\end{enumerate}
In our derivation, we ignore the cutoff contribution $\delta$. This contribution can be systematically included by repeated integration by parts if necessary. To approximate \eqref{IntSPMethod}, we expand $f(\vartheta')$ and $g(\vartheta')$ in power series around $\vartheta_*$:
\begin{equation}
    \begin{aligned}
        f(\vartheta') \approx f_0 + f_1\qty|\vartheta' - \vartheta_*|^{\ell} +f_2\qty|\vartheta' - \vartheta_*|^{\ell+k} +\mathcal{O}\qty[\qty|\vartheta' - \vartheta_*|^{\ell+2k}], \\
        g(\vartheta') \approx g_1\qty|\vartheta' - \vartheta_*|^{n-1} + g_2\qty|\vartheta' - \vartheta_*|^{n+m-1}+\mathcal{O}\qty[\qty|\vartheta' - \vartheta_*|^{n+2m-1}].
    \end{aligned} \label{SeriesExpansion}
\end{equation}
For simplicity, we assume that both $2k$ and $2m$ exceed $k$ and $m$ (\textit{i.e.}\ $2k>m$ and $2m>k$), so that the preceding terms are the largest contributions to the expansion. If this condition is not satisfied, the same method applies, but the series expansion must be truncated accordingly. Substituting \eqref{SeriesExpansion} into \eqref{IntSPMethod}, we perform the change of variables $t' = f_1 \lambda \abs{\vartheta' - \vartheta_*}^\ell$ (and similarly for $t''$), and expand the exponential up to order $\lambda^{-\frac*{k}{\ell}}$. The resulting integrals can then be evaluated using
\begin{equation}
    \begin{gathered}
         \int_0^{t(\delta)} \dd{t'} {t'}^{a - 1} e^{it'} \simeq \Gamma\qty(a)e^{\frac{i\pi a}{2}}, \\
         \int_0^{t(\delta)} \dd{t'} \int_0^{t'} \dd{t''} {t'}^{a - 1} {t''}^{b - 1} e^{i(t' - t'')} \simeq \frac{\Gamma\qty(a)\Gamma\qty(b) \sin\qty(\pi a)e^{\frac{i\pi \qty(a+b)}{2}}}{\sin\qty[\pi\qty(a+b)]},
    \end{gathered} \label{IndCutoffContributionInt}
\end{equation}
where $\simeq$ denotes the contribution independent of the cutoff $t(\delta)$. This yields the following approximations for \eqref{IntSPMethod}:
\begingroup\makeatletter\def\f@size{10}\check@mathfonts\def\maketag@@@#1{\hbox{\m@th\large\normalfont#1}}%
\begin{equation}
    \label{GeneralStationnaryPhase}
\begin{aligned}
    \mathcal{I}_1(\vartheta_*) &\simeq   \frac{g_1 e^{if_0 \lambda+\frac{i\pi n}{2\ell}}}{\ell \qty(f_1 \lambda )^{\frac{n}{\ell}}}  \left[ \Gamma\qty(\frac{n}{\ell})+ \frac{g_2}{g_1}\frac{\Gamma\qty(\frac{n+m}{\ell})e^{\frac{i\pi m}{2\ell}}}{\qty(f_1 \lambda )^{\frac{m}{\ell}}}  - \frac{f_2}{f_1}\frac{\Gamma\qty(\frac{n+ k}{\ell}+1)e^{\frac{i\pi k}{2\ell}}}{\qty(f_1 \lambda )^{\frac{k}{\ell}}}\right]+ \mathcal{O}\qty[\lambda^{-\frac{n+2\min\qty{k,m}}{\ell}}], \\
       \mathcal{I}_2(\vartheta_*) &\simeq \frac{g_1^2 e^{\frac{i n \pi}{\ell}}}{\ell^2(f_1\lambda)^{\frac{2n}{\ell}}} \left[ \frac{\Gamma\qty(\frac{n}{\ell})^2}{2\cos\qty(\frac{n\pi}{\ell})}  + \frac{g_2}{g_1}\frac{\Gamma \qty(\frac{n}{\ell })\Gamma \qty(\frac{n+m}{\ell})\cos \qty(\frac{m \pi }{2 \ell })  e^{\frac{i m \pi }{2\ell}}}{\qty(f_1 \lambda )^{\frac{m}{\ell}} \cos \qty( \frac{n \pi }{\ell} + \frac{m \pi }{2\ell})}      - \frac{f_2}{f_1}\frac{ \Gamma \qty(\frac{n}{\ell }) \Gamma \qty(\frac{n+k}{\ell}+1)\cos \qty(\frac{k \pi  }{2 \ell })  e^{\frac{ik\pi}{2\ell}}}{\qty(f_1\lambda)^{ \frac{k}{\ell} }\cos \qty(\frac{n \pi}{\ell} + \frac{k \pi}{2 \ell })}  \right] \\
    &\hspace{315pt}+ \mathcal{O}\qty[\lambda^{-\frac{2(n+\min\qty{k,m})}{\ell}}],
\end{aligned}
\end{equation}
\endgroup
where $\simeq$ again denotes the cutoff-independent terms, or equivalently, the contribution of the critical point. We note that \eqref{GeneralStationnaryPhase} diverges for certain values of $n, m, \ell, k$, reflecting the fact that the integrals in \eqref{IndCutoffContributionInt} are not well-defined for specific values of $a$ and $b$. Since such cases do not arise here, they are not considered further, and \eqref{GeneralStationnaryPhase} is assumed to hold whenever no divergence occurs.


\section{Magnetar Parameters} \label{Appendix:MagnetarsParameters}

This appendix summarizes the neutron star parameters used in our analysis to compute the bounds shown in Figures~\ref{fig:LowerBounds} and \ref{fig:UpperBounds}, see Table \ref{tab:MagnetarParameters}. Parameters listed without explicit references are taken from the \href{https://www.physics.mcgill.ca/~pulsar/magnetar/main.html}{McGill Online Magnetar Catalog} \cite{Olausen:2013bpa}. The original references for the $\nu F_\nu$ X-ray spectra used in this analysis are listed alongside each magnetar's name and the spectra are also tabulated in \cite{Fortin:2021sst}. The corresponding $\nu F_\nu$ spectra are compiled from observations performed by the Suzaku, INTEGRAL, NuSTAR, and RXTE missions.
\begin{table}[H]
    \centering
    \begin{tabular}{c|ccc}
    \hline\hline
    Magnetars & $B_0\, \qty[10^{14}\,\text{G}]$ & $d\, \qty[\text{kpc}]$    \\ \hline \hline
        1E 1547.0-5408$^{\text{\cite{Kuiper:2008, Kuiper:2012,CotiZelati2020}}}$ & 3.2  & 4.5 \\
        1E 1841-045$^{\text{\cite{Kuiper:2004ya, MoriiSuzaku:2010, Enoto:2017dox}}}$ & 7.0 & 8.5  \\
        1E 2259+586$^{\text{\cite{Enoto:2017dox} }}$& 0.59  & 3.2  \\
        4U 0142+61$^{\text{\cite{EnotoSuzaku:2011, Wang2013HardXE, Enoto:2017dox} }}$  &1.3 & 3.6 \\
        SGR 1806-20$^{\text{\cite{Enoto:2017dox, YounesBaringAl:2017}}}$  &7.7$^{\text{\cite{YounesBaringAl:2017}}}$ & 8.7$^{\text{\cite{Bibby2008ADR}}}$   \\
        \hline \hline
    \end{tabular}
    \caption{Magnetar parameters used in Figures~\ref{fig:LowerBounds} and \ref{fig:UpperBounds}. In all cases, the neutron star radius is taken to be $r_0 = 10\, \text{km}$ \cite{Kaspi:2017fwg}.}
    \label{tab:MagnetarParameters}
\end{table}


\section{Effective Cross Section} \label{Appendix:cte}

To evaluate the effective cross section of the conversion region \eqref{CrossSectionFluxRelation} analytically, we split the integration over $b$ into the two regimes $b\sim r_0$ and $b\gg r_0$, using the approximations \eqref{ParallelApproxbsimr0} and \eqref{LargebParallelApprox} in their respective domains. The matching point is set by the critical impact parameter $b_c$ defined in \eqref{critical_impact_parameter}, such that
\begin{align}
    \sigma_{\text{eff}} &\approx \frac{1}{2}\sum_j \int_0^{2\pi} \dd{\varphi}\qty{\int_{r_0}^{b_c} b\dd{b}  \left[P_{h_j \to \gamma_j}\right|_{\scalebox{0.67}{$\mqty{\hspace{-4pt}\alpha = 0 \vspace{-5pt}\\ \hspace{1pt} b \sim r_0}$}}
     + \int_{b_c}^{\infty} b\dd{b}  \left[P_{h_j \to \gamma_j}\right|_{\scalebox{0.67}{$\mqty{\hspace{-4pt}\alpha = 0 \vspace{-5pt}\\ \hspace{1pt} b \gg r_0}$}}
     }. \label{SectionEfficaceInteraction}
\end{align}
The contribution from the region $b\gg r_0$ is straightforward to evaluate using \eqref{LargebParallelApprox}, yielding
\begin{align}
   \int_{b_c}^{\infty} b\dd{b}  \left[P_{h_j \to \gamma_j}\right|_{\scalebox{0.67}{$\mqty{\hspace{-4pt}\alpha = 0 \vspace{-5pt}\\ \hspace{1pt} b \gg r_0}$}} &= \qty(\frac{65\,536}{17\,325})^2\frac{2\cdot 6^{\frac{1}{5}}(\Delta_{M0}r_0)^2r_0^2}{9\pi^2\qty(5\pi)^{\frac{2}{5}}(\Delta_{j 0 }r_0)^{\frac{2}{5}}}. \label{bggerIntegb}
\end{align}
In the region $b\sim r_0$, the integral is evaluated approximately using \eqref{ParallelApproxbsimr0}. After changing the integration variable to $\lambda$ [see \eqref{fgthetalambdaDefinition}], we expand the absolute value of \eqref{ParallelApproxbsimr0} up to order $\lambda^{-\frac{4}{3}}$. All of the remaining integrals over $\lambda$ will then adopt the following generic form
\begin{align}
    I(\chi, \nu, \zeta) &= \int_{1}^{\frac{45\pi \Delta_{j0}r_0}{256}} \dd{\lambda} \frac{\cos\qty(\nu \lambda + \zeta)}{\lambda^{{1-\chi}}}, \nonumber \\
    &\approx \nu^{-\chi} \Re\qty[\Gamma(\chi, -i\nu) e^{i\zeta +\frac{i\pi}{2}\chi} ], \label{ApproximationIntegCrossSection}
\end{align}
which is evaluated by extending the upper bound to infinity, justified by the fact that $\Delta_{j0}r_0$ is a large dimensionless parameter (see Section~\ref{sc:geometries}). This leads to the following approximation for the contribution from $b\sim r_0$ in \eqref{SectionEfficaceInteraction},
\begin{align}
    \int_{r_0}^{b_c} b\dd{b}  \left[P_{h_j \to \gamma_j}\right|_{\scalebox{0.67}{$\mqty{\hspace{-4pt}\alpha = 0 \vspace{-5pt}\\ \hspace{1pt} b \sim r_0}$}} &\approx C\cdot \frac{(\Delta_{M0}r_0)^2r_0^2}{(\Delta_{j 0 }r_0)^{\frac{2}{5}}}, \label{bsimr0Integbapprox}
\end{align}
where the numerical constant $C = 0.89274\ldots$ admits an exact representation in terms of incomplete gamma and trigonometric functions given by
\begin{align*}
    C &=\frac{(5 \pi )^{\frac{26}{35}} \Gamma\qty(\frac{3}{7}) \Gamma \qty(\frac{5}{7}) \left\{35\cdot
   2^{\frac{9}{35}} \cos\qty(\frac{\pi }{7})+52\Re\left[\qty(-1)^{\frac{2}{35}} \Gamma \qty(-\frac{26}{35},-2 i)\right]\right\}}{832 \cdot 2^{\frac{1}{5}} 3^{\frac{4}{5}} 7^{\frac{6}{7}}}\\
   &\hspace{50pt}+\frac{3 \cdot 3^{\frac{1}{5}} (5 \pi )^{\frac{16}{35}} \Gamma \qty(\frac{3}{7})^2 \left\{35+16\cdot 2^{\frac{16}{35}} \Re\left[\qty(-1)^{\frac{12}{35}} \Gamma \qty(-\frac{16}{35},-2
   i)\right]\right\}}{2240\cdot 2^{\frac{23}{35}} 7^{\frac{1}{7}}} \\
   &\hspace{100pt}+\frac{3\cdot 3^{\frac{11}{30}} \qty(10
   \pi )^{\frac{73}{105}}  \Gamma \qty(\frac{2}{3}) \Gamma \qty(\frac{3}{7}) \Re\left[\qty(-1)^{\frac{197}{210}} \Gamma \qty(-\frac{73}{105},-i)\right]}{80\cdot 2^{\frac{9}{35}}
   7^{\frac{4}{7}}}\\
   &\hspace{150pt} +  \frac{27 (5\pi)^{\frac{14}{15}} \Gamma\qty(\frac{2}{3})^2}{1792\cdot 6^{\frac{7}{15}}}, \\
   &\approx  0.89274.
\end{align*}
Combining \eqref{bggerIntegb} and \eqref{bsimr0Integbapprox} yields the approximation \eqref{ApproximateEffectiveCrossSectionDeltas} for the effective cross section.

\newpage
\bibliography{Gertsenshtein}


\end{document}